\documentclass[a4paper,USenglish, cleveref, autoref]{lipics-v2019}

\usepackage[nolist,nohyperlinks]{acronym}
\usepackage{booktabs}
\usepackage{xfrac}
\usepackage{url}

\usepackage[all]{background}
\backgroundsetup{color=black!20!blue, opacity={1.0}, angle={0}, anchor={right}, scale={0.9}, position={2cm, -0.1cm}}
\SetBgContents{\parbox{1.4\textwidth}{ \centering
					This is a pre-print. The final version of the article is published as part of the proceedings of the 31st Euromicro Conference \\ on Real-Time Systems (ECRTS 2019) and is available online at \url{https://doi.org/10.4230/LIPIcs.ECRTS.2019.12}.
		     	 }
	     	 }

\makeatletter
\newcommand*{\org@overidelabel}{}
\let\org@overridelabel\@verridelabel
\@ifpackagelater{acronym}{2015/03/21}{% v1.41
  \renewcommand*{\@verridelabel}[1]{%
    \@bsphack
    \protected@write\@auxout{}{\string\AC@undonewlabel{#1@cref}}%
    \org@overridelabel{#1}%
    \@esphack
  }%
}{% older versions
  \renewcommand*{\@verridelabel}[1]{%
    \@bsphack
    \protected@write\@auxout{}{\string\undonewlabel{#1@cref}}%
    \org@overridelabel{#1}%
    \@esphack
  }%
}
\makeatother

\title{Isolation-Aware Timing Analysis and Design Space Exploration for Predictable and Composable Many-Core Systems}

\titlerunning{Isolation-Aware Timing Analysis and Design Space Exploration for Many-Cores}

\author{Behnaz Pourmohseni}{Friedrich-Alexander-Universit\"at Erlangen-N\"urnberg (FAU), Germany}{behnaz.pourmohseni@fau.de}{https://orcid.org/0000-0003-0350-4784}{}

\author{Fedor Smirnov}{Friedrich-Alexander-Universit\"at Erlangen-N\"urnberg (FAU), Germany}{fedor.smirnov@fau.de}{}{}

\author{Stefan Wildermann}{Friedrich-Alexander-Universit\"at Erlangen-N\"urnberg (FAU), Germany}{stefan.wildermann@fau.de}{}{}

\author{J\"urgen Teich}{Friedrich-Alexander-Universit\"at Erlangen-N\"urnberg (FAU), Germany}{juergen.teich@fau.de}{}{}

\authorrunning{B.\ Pourmohseni, F.\ Smirnov, S.\ Wildermann, and J.\ Teich}

\Copyright{Behnaz Pourmohseni, Fedor Smirnov, Stefan Wildermann, and J\"urgen Teich}

\ccsdesc[500]{Computer systems organization~Real-time systems}
\ccsdesc[300]{Computer systems organization~Embedded and cyber-physical systems}
\ccsdesc[500]{Computer systems organization~Multicore architectures}

\keywords{Many-core systems, timing analysis, design space exploration (DSE), isolation scheme, predictability, composability.}

\acknowledgements{This work is funded by the Deutsche Forschungsgemeinschaft (DFG, German Research Foundation) - Project Number 146371743 - TRR 89 Invasive Computing.}

\nolinenumbers

\EventEditors{Sophie Quinton}
\EventNoEds{1}
\EventLongTitle{The final article is available in the proc.\ of the 31st Euromicro Conference on Real-Time Systems (ECRTS 2019)}
\EventShortTitle{ECRTS 2019}
\EventAcronym{ECRTS}
\EventYear{2019}
\EventDate{July 9--12, 2019}
\EventLocation{Stuttgart, Germany}
\EventLogo{}
\SeriesVolume{133}
\ArticleNo{12}
\begin{document}

\maketitle

\begin{abstract}
Composable many-core systems enable the independent development and analysis of applications which will be executed on a shared platform where the mix of concurrently executed applications may change dynamically at run time.
For each individual application, an off-line \ac{DSE} is performed to compute several mapping alternatives on the platform, offering Pareto-optimal trade-offs in terms of real-time guarantees, resource usage, etc.
At run time, one mapping is then chosen to launch the application on demand.
In this context, to enable an independent analysis of each individual application at design time, so-called \emph{inter-application isolation schemes} are applied which specify temporal/spatial isolation policies between applications.
State-of-the-art composable many-core systems are developed based on a \emph{fixed} isolation scheme that is exclusively applied to every resource in every mapping of every application and use a timing analysis \emph{tailored} to that isolation scheme to derive timing guarantees for each mapping.
A \emph{fixed} isolation scheme, however, heavily restricts the explored space of solutions and can, therefore, lead to suboptimality.
Lifting this restriction necessitates a timing analysis that is applicable to mappings with an arbitrary mix of isolation schemes on different resources.
To address this issue, in this paper, we (a)~present an \emph{isolation-aware timing analysis} that---unlike existing analyses---can handle multiple isolation schemes in combination within one mapping and delivers safe yet tight timing bounds by identifying and excluding interference scenarios that can never happen under the given combination of isolation schemes.
Based on the timing analysis, we (b)~present a \ac{DSE} which explores the choices of isolation scheme per resource within each mapping and uses the proposed timing analysis for timing verification.
Experimental results demonstrate that, for a variety of real-time applications and many-core platforms, the proposed approach achieves an improvement of up to 67\% in the quality of delivered mappings compared to approaches based on a fixed isolation scheme.
\end{abstract}

%%%%%%%%%%%%%%%%%%%%%%%%%%%%%%%%%%%%%%%%%%%%%%%%%%%%%%
%%%%%%%%%%%%%%%%%%%%%%%%%%%%%%%%%%%%%%%%%%%%%%%%%%%%%% DEFINITIONS
%%%%%%%%%%%%%%%%%%%%%%%%%%%%%%%%%%%%%%%%%%%%%%%%%%%%%%

%%%%%%%%%%%%%%%%%%%%%%%%%%%%%%%%%%%%%%%%%%%%%%%%% acronyms
\acrodef{DSE}[DSE]{Design Space Exploration}
\acrodef{NOC}[NoC]{Network-on-Chip}
\acrodef{HAM}[HAM]{Hybrid Application Mapping}
\acrodef{TDM}[TDM]{Time-Division Multiplexing}
\acrodef{WCTT}[WCTT]{Worst-Case Transition Time}
\acrodef{WCET}[WCET]{Worst-Case Execution Time}
\acrodef{WCRT}[WCRT]{Worst-Case Response Time}
\acrodef{WCTT}[WCTT]{Worst-Case Traversal Time}
\acrodef{NORMA}[NORMA]{No Remote Memory Access}
\acrodef{MPSoC}[MPSoC]{Multi-Processor System-on-Chip}
\acrodef{E3S}[E3S]{Embedded System Synthesis Benchmarks Suite}
\acrodef{NA}[NA]{Network Adapter}
\acrodef{WRR}[WRR]{Weighted Round-Robin}
\acrodef{RR}[RR]{Round-Robin}
\acrodef{TX}[TX]{Transmitter}
\acrodef{RX}[RX]{Receiver}
\acrodef{FP}[FP]{Fixed-Priority}
\acrodef{FCFS}[FCFS]{First Come First Serve}

%%%%%%%%%%%%%%%%%%%%%%%%%%%%%%%%%%%%%%%%%%%%%%%%% symbols
\newcommand{\symMapping}{v}
\newcommand{\symAllMappings}{V}
\newcommand{\symTask}{t}
\newcommand{\symAllTasks}{T}
\newcommand{\symMessage}{m}
\newcommand{\symAllMessages}{M}
\newcommand{\symAppGraphEdge}{e}
\newcommand{\symAppGraphAllEdges}{E}
\newcommand{\symTaskPeriod}{\text{PRD}}
\newcommand{\symMessagePeriod}{\text{PRD}}
\newcommand{\symTaskWCET}{\text{WCET}}
\newcommand{\symTaskMD}{\text{MD}}
\newcommand{\symMessageMD}{\text{MD}}
\newcommand{\symMessageSize}{\text{PLD}}

\newcommand{\symProcessor}{c}
\newcommand{\symAllProcessors}{C}
\newcommand{\symRouter}{r}
\newcommand{\symAllRouters}{R}
\newcommand{\symBus}{b}
\newcommand{\symAllBuses}{B}
\newcommand{\symNI}{n}
\newcommand{\symAllNIs}{N}
\newcommand{\symTX}{\text{\textit{tx}}}
\newcommand{\symRX}{\text{\textit{rx}}}
\newcommand{\symTile}{u}
\newcommand{\symAllTiles}{U}
\newcommand{\symMemory}{q}
\newcommand{\symAllMemories}{Q}
\newcommand{\symLink}{l}
\newcommand{\symAllLinks}{L}
\newcommand{\symNocClock}{\tau^\text{noc}}
\newcommand{\symMemDelay}{\text{ST}}

\newcommand{\symArbPeriod}{P}
\newcommand{\symArbInterval}{S}
\newcommand{\symArbWeight}{W}
\newcommand{\symArbDelay}{D}
\newcommand{\symArbCapacity}{K}

%%%%%%%%%%%%%%%%%%%%%%%%%%%%%%%%%%%%%%%%%%%%%%%%%%%%%%
%%%%%%%%%%%%%%%%%%%%%%%%%%%%%%%%%%%%%%%%%%%%%%%%%%%%%% (1) INTORDUCTION
%%%%%%%%%%%%%%%%%%%%%%%%%%%%%%%%%%%%%%%%%%%%%%%%%%%%%%
\section{Introduction}
\begin{figure}[!t]
\centering
\includegraphics[width=0.9\linewidth,keepaspectratio]{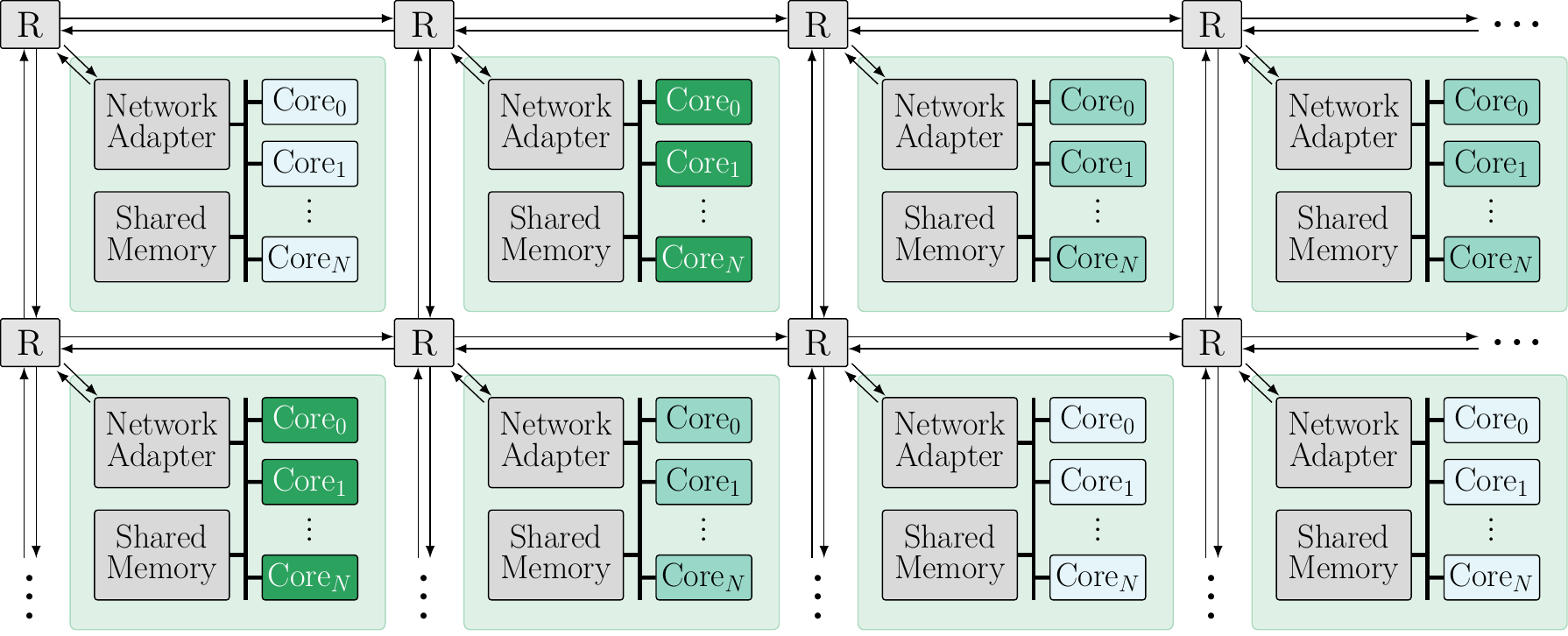}
\caption{An example of a tiled many-core architecture composed of heterogeneous compute tiles.}
\label{fig:manyCore}
\end{figure}

\acresetall
The ever-growing number of applications hosted in modern embedded systems introduces a high compute power demand which has given rise to the recent shift towards many-core architectures, e.g., Tilera TILE-Gx~\cite{tile2012tilera}, Kalray MPPA-256~\cite{de2013clustered}, and Intel SCC~\cite{howard2010a}.
For scalability, a many-core architecture is often organized as a set of compute tiles interconnected via a \ac{NOC}, see~\cref{fig:manyCore}.
Each compute tile comprises a \ac{NA}, a set of processing cores and peripherals such as memories which are interconnected via a set of buses.
From a designer's perspective, the large number and diversity of applications and their non-functional requirements, e.g., real-time constraints, in modern embedded systems introduces an immense system design complexity.
This renders the \emph{integrated} system design approach, in which the whole system is designed at once, impractical. 
Over the past decade, \emph{composable} systems, e.g.~\cite{hansson2009compsoc}, have emerged to address this issue.
In a composable system, applications are temporally and/or spatially isolated from each other, enabling an \emph{incremental} system design approach where each application is first developed and analyzed individually and then added to the system on demand~\cite{akesson2011composability}.

Composability is particularly crucial for the development of dynamic many-core systems with variable workloads and hard real-time requirements.
In such systems, several independent applications are executed simultaneously, each being launched and terminated on demand and independently from others, resulting in a dynamic mix of active applications and, thus, a dynamic availability of platform resources.
Each running application may be exposed to a variable workload which corresponds to a dynamic compute power demand to, e.g., meet real-time constraints.
Moreover, unforeseeable conditions, e.g., thermal hot spots and hardware faults, affect the execution of applications that are currently running on the affected regions of the platform.
Reserving resources according to the worst-possible workload scenario often leads to an immense underutilization of system resources which is typically not acceptable due to cost concerns.
To cope with such dynamics in both resource availability of the system and resource demand of applications, \emph{\ac{HAM}} strategies have emerged recently~\cite{singh2013mapping, weichslgartner2018dissertation}.
In \ac{HAM}, each application is developed and analyzed individually using an off-line \ac{DSE} which computes several deployment options, so-called \emph{mappings}, of the application on the platform.
The computed mappings are ensured to offer diverse resource demand and performance guarantees to address various run-time resource-availability and workload scenarios, respectively.
The mappings computed for each application are then provided to a so-called \emph{run-time platform manager} which launches each application on demand using a precomputed mapping that satisfies the on-line performance requirements of the application and the resource constraints of the platform.
Moreover, if the mapping in use by a running application fails, e.g., due to thermal hot spots, resource faults, or a drastic workload change, the run-time manager switches the application to another precomputed mapping which conforms to the new conditions~\cite{pourmohseni2019hard}.

\begin{figure}[!t]
\centering
\includegraphics[width=0.95\linewidth,keepaspectratio]{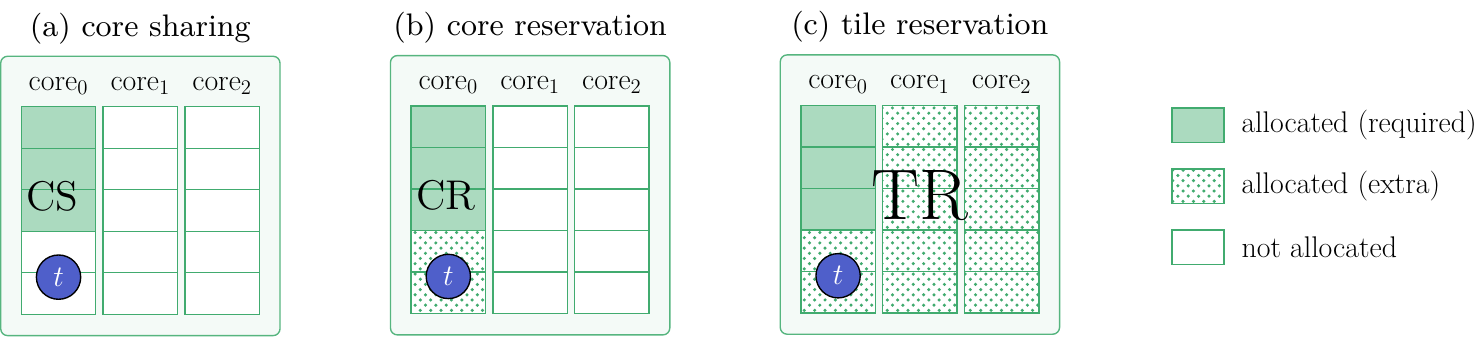}
\caption{An example of a task $\symTask$ mapped to core\textsubscript{0} of a 3-core tile under (a)~core-sharing, (b)~core-reservation, and (c)~tile-reservation isolation schemes. For a timely completion, $\symTask$ requires a core budget of 3 (shaded cells). Dotted cells are allocated in addition due to the isolation scheme in use.}
\label{fig:motivationIndividualSchemes}
\end{figure}

The off-line \ac{DSE} used in \ac{HAM} strategies employs compute-intensive optimization and verification techniques to find high-quality mappings that offer Pareto-optimal trade-offs w.r.t.\ multiple---oftentimes non-linear and conflicting---design objectives, e.g., resource usage, latency, energy, etc. 
To achieve \emph{predictability}, the worst-case timing properties of each mapping are bounded based on the choice of allocated resources by deriving the worst-case timing interferences that may be imposed by other (concurrent) applications which share resources with the mapping under analysis~\cite{weichslgartner2014daarm}. 
During the \ac{DSE}, however, an application is developed individually, and the characteristics of the potential concurrent applications are not available to be considered in the timing analysis.
In order to regulate the maximum degree of timing interferences that may be imposed by other applications, temporal and spatial isolation techniques are employed which apply certain restrictions on the accessibility of resources used by the mapping under analysis to other applications that run concurrently.
These restrictions are referred to as so-called \emph{inter-application isolation schemes}.
\Cref{fig:motivationIndividualSchemes} illustrates three major isolation schemes that are predominantly used in many-core systems to establish composability.
In all three cases, an exemplary task $\symTask$ is mapped to core\textsubscript{0} of a 3-core tile.
For the sake of brevity, only cores are depicted in the illustration of the tile.
The compute power of each core is divided into 5 budgets of equal size. 
For a timely completion, $\symTask$ requires a budget of 3 on core\textsubscript{0}.
In the following, each isolation scheme is explained.
\begin{itemize}
  \item \textbf{Core Sharing (CS).} In the isolation scheme referred to as core sharing, illustrated in \cref{fig:motivationIndividualSchemes}a, only the 3 core budgets required for $\symTask$ are allocated for the mapping.
	Hence, concurrent applications are \emph{temporally isolated} from the current mapping on core\textsubscript{0} and can use the 2 not-allocated budgets of core\textsubscript{0}, the whole budget of core\textsubscript{1} and core\textsubscript{2}, and any other on-tile resource, e.g., memories and the \ac{NA}.
	Core sharing is the least restrictive isolation scheme which merely depends on temporal isolation on all resources.
  \item \textbf{Core Reservation (CR).} In the isolation scheme referred to as core reservation, illustrated in \cref{fig:motivationIndividualSchemes}b, core\textsubscript{0} is allocated as a whole, regardless of the budget demand of $\symTask$.
  This realizes a \emph{spatial isolation} from other applications on core\textsubscript{0}, eliminating interferences from them on core\textsubscript{0} and resulting in an alleviated worst-case latency compared to core sharing.
  Note that interferences may still arise on other on-tile resources, e.g., memory buses used by $\symTask$, as applications are temporally isolated from each other on those resources.
  \item \textbf{Tile Reservation (TR).} In the isolation scheme referred to as tile reservation, illustrated in \cref{fig:motivationIndividualSchemes}c, the compute tile is allocated as a whole, eliminating any interference from concurrent applications on any on-tile resource.
  This is the most restrictive isolation scheme which realizes a \emph{spatial isolation} from concurrent applications at tile level and enables the largest reduction in the worst-case timing interferences imposed on $\symTask$.
\end{itemize}	

\begin{figure}[!t]
\centering
\includegraphics[width=0.99\linewidth,keepaspectratio]{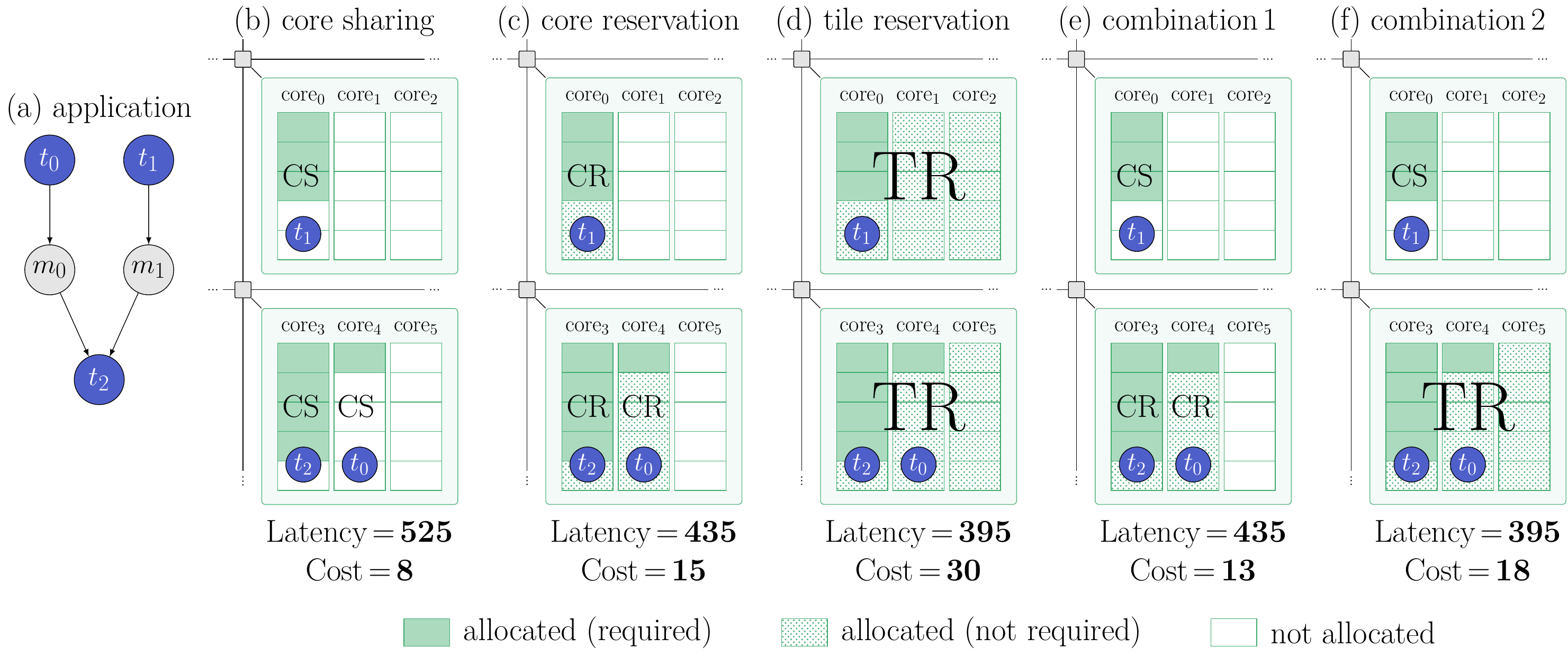}
\caption{
(a) an example of an application. A mapping of the application on two adjacent tiles is illustrated for the following isolation scenarios: (b) exclusive core sharing, (c) exclusive core reservation, (d) exclusive tile reservation, (e) core sharing on core\textsubscript{0} and core reservation on core\textsubscript{3} and core\textsubscript{4}, (f) core sharing on core\textsubscript{0} and tile reservation on the lower tile.
The core budget required for each task is shaded on the respective core.
For each scenario, the worst-case latency of the mapping (derived based on~\cref{tab:tabTiming}a--b) and its allocation cost are given below the respective sub-figure.
}
\label{fig:motivationIsolationSchemes}
\end{figure}

\begin{table}[b]
  \centering
  \caption{Qualitative timing properties assumed for tasks and messages in the example illustrated in~\cref{fig:motivationIsolationSchemes}: (a) WCRT of tasks $\symTask_0$, $\symTask_1$, and $\symTask_2$ under different isolation schemes, (b) WCTT of message $\symMessage_1$ for different combinations of isolation scheme on $\symMessage_1$'s source and destination cores/tiles.}
  \label{tab:tabTiming}
  \resizebox{1\columnwidth}{!}{
  \begin{tabular}{cccc}
   \multicolumn{4}{c}{(a) Worst-Case Response Time assumed for $\symTask_{0}$--$\symTask_{2}$} \\[0.4em]
   \toprule
    isolation & \multirow{ 2}{*}{WCRT($\symTask_0$)} & \multirow{ 2}{*}{WCRT($\symTask_1$)} & \multirow{ 2}{*}{WCRT($\symTask_2$)} \\ 
    scheme & & & \\[0.25em]
   \midrule
    TR & $290$ & $75$ & $105$ \\[0.3em]
    CR & $315$ & $115$ & $120$ \\[0.3em]
    CS & $380$ & $165$ & $145$ \\[0.15em]
   \bottomrule
  \end{tabular}
  \hspace{0.9 cm}
  \begin{tabular}{ccc}
   \multicolumn{3}{c}{(b) Worst-Case Traversal Time assumed for $\symMessage_1$ } \\[0.4em]
   \toprule
    source & destination & \multirow{ 2}{*}{WCTT($\symMessage_1$)} \\ 
    isolation scheme & isolation\ scheme & \\ 
   \midrule
    TR & TR & $12$ \\
    TR & CR/CS & $15$ \\
    CR/CS & TR & $17$ \\
    CR/CS & CR/CS & $20$ \\
   \bottomrule
  \end{tabular}
  }
 \end{table}

Noteworthy, sharing the \ac{NOC} which interconnects the tiles can hardly be avoided~\cite{mitra2018time}.
Hence, applications are always \emph{temporally isolated} from each other on the \ac{NOC}.
State-of-the-art composable many-core systems are designed based on a \emph{fixed isolation scheme} that is uniformly applied to \emph{every} core/tile of \emph{every} mapping of \emph{every} application in the system and is coupled with a timing analysis \emph{tailored} to that specific isolation scheme to derive tight timing guarantees.
The choice of isolation scheme highly impacts the resource usage and worst-case timing characteristics of the mappings.
As an example, consider the application shown in~\cref{fig:motivationIsolationSchemes}a and a mapping of it on two adjacent 3-core tiles illustrated in~\cref{fig:motivationIsolationSchemes}b--f.
For the sake of this motivational example, assume that the timing analysis of the tasks yields the qualitative \ac{WCRT} values given in~\cref{tab:tabTiming}a under different isolation schemes.
Likewise, the qualitative \ac{WCTT} of message $\symMessage_{1}$ (communicated between the two tiles, from $\symTask_1$ to $\symTask_2$) is given in~\cref{tab:tabTiming}b for various combinations of isolation scheme on $\symMessage_{1}$'s source and destination cores/tiles.
Message $\symMessage_0$ is implicitly communicated between $\symTask_0$ and $\symTask_2$ via the tile's shared memories.
In this example, applying core sharing exclusively, as illustrated in~\cref{fig:motivationIsolationSchemes}b, offers the minimum allocation cost, i.e., 8 core budgets, at the expense of considerably high worst-case interferences on cores and other on-tile resources, resulting in an end-to-end application latency of 525 time units in the worst case.
Applying core reservation exclusively, as depicted in~\cref{fig:motivationIsolationSchemes}c, results in an elevated allocation cost of 15 budgets.
It, however, eliminates core interferences which alleviates the worst-case application latency to 435 time units.
Applying tile reservation exclusively, as depicted in~\cref{fig:motivationIsolationSchemes}d, minimizes the worst-case application latency to 395 time units at the expense of a maximized allocation cost of 30 budgets.

 \begin{figure}[!t]
\centering
\includegraphics[width=0.55\linewidth,keepaspectratio]{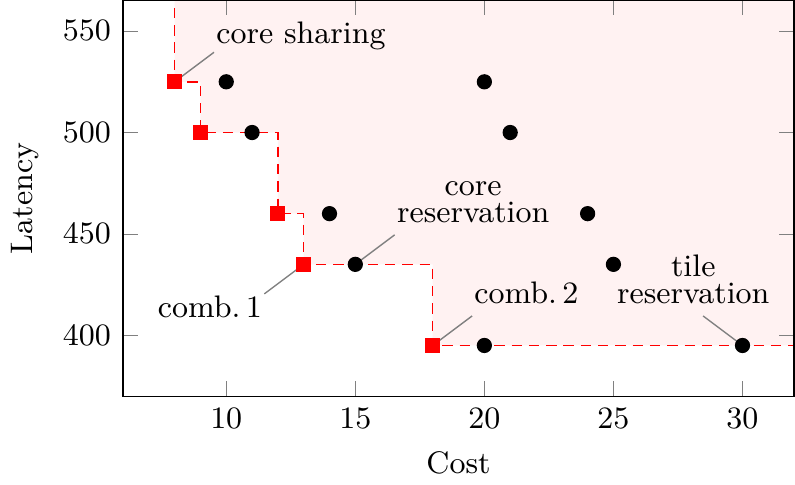}
\caption{The latency-cost trade-offs offered by the 15 possible isolation scheme scenarios for the application mapping from~\cref{fig:motivationIsolationSchemes}. Pareto-optimal trade-offs are designated by red squares, dominating the highlighted space. Labeled trade-offs correspond to the scenarios illustrated in~\cref{fig:motivationIsolationSchemes}.}
\label{fig:motivationFront}
\end{figure}

\textbf{Motivation.}
A \emph{fixed isolation scheme} that is always applied exclusively---which is the common practice in state-of-the-art composable many-core systems---heavily restricts the space of explored mappings and excludes numerous promising solutions where multiple isolation schemes are applied \emph{in combination}.
For instance, in the example above, considering arbitrary combinations of isolation schemes within one mapping enables 12 additional isolation scenarios, two of which are depicted in~\cref{fig:motivationIsolationSchemes}e--f.
In combination\,1, core sharing is applied on core\textsubscript{0} and core reservation is applied on core\textsubscript{3} and core\textsubscript{4}, offering the same latency as the exclusive core-reservation scenario from~\cref{fig:motivationIsolationSchemes}c, but at a lower cost.
Likewise, combination\,2 applies core sharing on core\textsubscript{0} and tile reservation on the lower tile, outperforming the exclusive tile-reservation scenario from~\cref{fig:motivationIsolationSchemes}d by offering the same latency at a considerably lower cost.
Having evaluated all possible combinations of isolation scenarios, \cref{fig:motivationFront} illustrates the obtained cost-latency trade-offs where Pareto-optimal trade-offs are designated by red squares, dominating the highlighted space above the red dashed line.
The trade-offs corresponding to the scenarios in~\cref{fig:motivationIsolationSchemes} are labeled accordingly in~\cref{fig:motivationFront}.
As shown, nearly always, Pareto-optimal trade-offs are obtained only when multiple isolation schemes are applied in combination within one mapping.
In~\cref{sec:results}, we experimentally verify this observation for realistic use cases as well.

\textbf{Contribution.}
The example above demonstrates that using multiple isolation schemes in combination yields mappings of a better quality trade-off.
Deriving safe bounds on the worst-case timing characteristics of such mappings, however, requires an \emph{isolation-aware timing analysis} that can capture the impact of the isolation scheme of each core/tile on the worst-case timing behavior of tasks/messages affected by it and, hence, on the worst-case timing behavior of the application.
On the one hand, existing timing analyses applicable to composable many-core systems are \emph{tailored} to one fixed isolation scheme and cannot handle multiple isolation schemes in combination within one mapping.
On the other hand, from an architectural point of view, many-core platforms are undergoing a transition from single-core tiles, e.g., in~\cite{tile2012tilera}, to multi-core tiles, e.g., in~\cite{de2013clustered}.
This transition introduces new sources of timing interference within each tile, e.g., on shared memory buses, which must be accounted for to obtain safe timing guarantees.
In the context of composable systems, there exist no timing analysis known to us which accounts for all sources of timing interference that may arise in a many-core platform with multi-core tiles (without applying restrictive spatial isolation schemes that eliminate on-tile inter-application interferences altogether). 
Provided with an isolation-aware timing analysis, the off-line \ac{DSE} in \ac{HAM} strategies can be extended to also explore the choices of isolation scheme for each core/tile within each investigated mapping to obtain solutions with a better quality trade-off.
In this line, the paper at hand makes the following contributions in the context of composable many-core systems:
\begin{enumerate}
 \item We present an \emph{isolation-aware timing analysis} which can handle arbitrary combinations of isolation schemes within one mapping and accounts for all timing interferences that may arise within a shared multi-core tile and on the \ac{NOC} interconnecting tiles. It derives tight timing bounds by identifying and excluding interference scenarios that can never happen under the current combination of isolation schemes.
 \item We present an \emph{isolation-aware \ac{DSE}} which explores the choices of isolation scheme for each core/tile in each investigated mapping of the application under analysis and uses the proposed timing analysis for the scheduling and timing verification of the mappings.
\end{enumerate}
For a variety of hard real-time applications and many-core platforms, we experimentally verify the advantage of the proposed isolation-aware exploration and timing analysis approach over existing fixed-isolation-scheme approaches in terms of the quality of obtained mappings.

\textbf{Organization.}
The remainder of this paper is organized as follows.
In \cref{sec:relatedwork}, related work on timing analysis and \ac{DSE} of multi-/many-core systems is reviewed.
\Cref{sec:preliminaries} presents the preliminaries and the system model for this work.
The proposed isolation-aware \ac{DSE} and timing analysis are presented in~\cref{sec:dse,sec:timingAnalysis}, respectively.
Experimental results are discussed in~\cref{sec:results} before the paper is concluded in~\cref{sec:conclusion}.

%%%%%%%%%%%%%%%%%%%%%%%%%%%%%%%%%%%%%%%%%%%%%%%%%%%%%%
%%%%%%%%%%%%%%%%%%%%%%%%%%%%%%%%%%%%%%%%%%%%%%%%%%%%%% (2) RELATED WORKS
%%%%%%%%%%%%%%%%%%%%%%%%%%%%%%%%%%%%%%%%%%%%%%%%%%%%%%

\section{Related Work}
\label{sec:relatedwork}

Worst-case timing analysis of applications in multi-/many-core systems has long been conducted in two steps: 
First, a context-independent analysis is applied to bound the \ac{WCET} of each task in isolation, i.e., in absence of interferences.
An overview of tools and methods for context-independent \ac{WCET} analysis is provided in~\cite{wilhelm2008worst}.
Following the context-independent analysis, an interference analysis is performed to bound the additional latencies that may be imposed on shared resources due to external interferences.

Multi-/many-core timing analyses predominantly focus on worst-case interference analysis based on the context-independent characteristics of each task and message.
For instance, the analyses presented in~\cite{altmeyer2015generic, dasari2016framework, davis2017extensible, giannopoulou2012timed, giannopoulou2016mixed, rihani2016response, rouxel2017tightening, skalistis2016worst} bound the \ac{WCRT} of tasks in a multi-core setup where several cores are connected to one or more memories over shared buses.
In~\cite{altmeyer2015generic, davis2017extensible}, a framework for multi-core response time analysis is presented for a preemptive \ac{FP} core scheduling policy coupled with multiple memory bus arbitration policies, e.g., \ac{TDM}, \ac{RR}, and \ac{FP}.
Targeting mixed-criticality systems,~\cite{giannopoulou2012timed, giannopoulou2016mixed} analyze \ac{WCRT} under a non-preemptive \ac{FP} core scheduling policy and a \ac{RR} memory bus arbitration.
The authors of~\cite{rihani2016response, skalistis2016worst} present response time analyses for non-preemptive \ac{FP} core scheduling policy and a multi-level bus arbitration scheme which combines \ac{RR} and \ac{FP} policies.
In~\cite{rouxel2017tightening}, memory bus interference is bounded using an ILP-based timing analysis under a \ac{RR} bus arbitration policy.
The framework presented in~\cite{dasari2016framework}, analyzes memory bus interference under a variety of arbitration policies, e.g., \ac{TDM} and \ac{FP}.
Concerning the \ac{NOC}, in~\cite{shi2008real} and~\cite{zhan2013designing} \ac{NOC} transfer delays are analyzed under \ac{FP} and \ac{RR} link arbitration policies, respectively.

The works listed above consider---to some extent---non-preemptive and/or \emph{contention-oriented} arbitration/scheduling policies, e.g., priority-based schemes, for which safe timing guarantees can be derived only if the whole set of applications accessing shared resources is known at the time of analysis.
In a composable system, however, the mix of concurrently running applications that share some resources may not be known at design time.
Hence, contention-oriented policies can be applied in a composable system only if each and every resource with such an arbitration scheme is exclusively used by one application.
To realize this on state-of-the-art many-core platforms, each mapping computed for each application must (a)~have a tile-reservation isolation scheme only, (b)~be restricted within one tile only, and (c)~not rely on inter-tile communications, e.g., I/O transfers.
In practice, however, the small number of cores comprised within one tile~\cite{tile2012tilera,howard2010a}, on the one hand, and the high compute power demand of each application, on the other hand, necessitate the deployment of some applications over several compute tiles to meet their real-time requirements.
For such applications, \ac{NOC} links could---in theory---be exclusively reserved per application to enable the timing analysis of multi-tile mappings using \ac{NOC} delay analyses, e.g.~\cite{zhan2013designing, shi2008real}.
In practice, however, sharing \ac{NOC} links among applications can hardly be avoided~\cite{mitra2018time}.

\emph{Contention-free} arbitration policies based on time slicing, e.g., \ac{TDM} and \ac{WRR}, offer a practical approach for predictable inter-application resource sharing without violating composability.
For instance, the authors of~\cite{weichslgartner2014daarm, weichslgartner2018design} consider a \ac{NOC}~\cite{heisswolf2013providing} with \ac{WRR} link arbitration policy and analyze worst-case \ac{NOC} delays based on the reserved link budget for each communication, independent of the other communication flows that may share the same links.
For \ac{WCRT} analysis,~\cite{weichslgartner2014daarm} considers a preemptive \ac{RR} core scheduling policy which, however, restricts its scope of coverage to core interferences originating from within the application under analysis only.
In~\cite{weichslgartner2018design}, on the other hand, a \ac{WRR} core scheduling policy is considered, and based on that, a response time analysis is presented which accounts also for core interferences that may be imposed by other (currently unknown) applications.
Both~\cite{weichslgartner2014daarm, weichslgartner2018design}, however, consider single-core tiles and, hence, cannot capture \ac{NA} and memory bus interferences that may arise in a \emph{multi-core tile}, e.g., in~\cite{de2013clustered, howard2010a}.
Contrarily, we present a timing analysis that captures also the \ac{NA} and memory bus interferences in multi-core tiles.

From an isolation scheme point of view, existing multi-/many-core timing analyses are tailored to a fixed isolation scheme.
For instance, the analyzes in~\cite{altmeyer2015generic, dasari2016framework, davis2017extensible, giannopoulou2012timed, giannopoulou2016mixed, rihani2016response, rouxel2017tightening, skalistis2016worst} are applicable only under tile-reservation isolation scheme.
The analysis presented in~\cite{weichslgartner2014daarm} is applicable to systems with single-core tiles only and assumes tile-reservation isolation scheme.
Authors in~\cite{weichslgartner2018design} consider single-core tiles and core-sharing isolation scheme.
The timing analysis presented in this paper is the first timing analysis known to us which can handle arbitrary mixes of isolation schemes in combination within one mapping.

Application mapping in multi-/many-core systems is typically viewed as a multi-objective optimization problem and is known to be NP-hard~\cite{garey1990guide}.
Due to its immensely large space of possible mapping solutions, using exact optimization approaches, e.g., enumeration or (integer-)linear programming, to solve this NP-hard problem demands an extremely high computational effort and is prohibitively time-consuming, except for very small/simple problems.
In this context, \emph{meta-heuristic optimization} approaches, e.g., evolutionary/genetic algorithms~\cite{davis1991handbook, fonseca1995overview, fonseca1993genetic}, simulated annealing~\cite{kirkpatrick1983optimization}, and particle swarm optimization~\cite{kennedy2010particle}, enable a scalable optimization approach that can deliver high-quality solutions at a reasonable time- and computational effort.
As a result, they have become the de facto standard approach for multi-objective application mapping optimization in multi-/many-core systems.
For instance, in~\cite{mariani2010industrial, piscitelli2012design, weichslgartner2014daarm, weichslgartner2018design}, evolutionary/genetic algorithms are used for mapping optimization.
In~\cite{giannopoulou2013scheduling, giannopoulou2016mixed}, mapping optimization is realized using simulated annealing algorithms, while~\cite{sahu2014application} adopts particle swarm optimization.
The majority of existing works on mapping optimization in \ac{HAM} methodologies employ \emph{population-based} meta-heuristics, e.g., particle swarm optimization or evolutionary/genetic algorithms, to collect a so-called \emph{population} of best mappings.
During the Design Space Exploration (DSE), the optimizer follows the course of several iterations to generate new mappings and evaluate them w.r.t.\ the given set of design objectives, e.g., resource cost, timing, or energy.
It collects a population of explored mappings which is updated per iteration to retain the best solutions found so far and is used for the generation of mappings in the next iteration.
Like~\cite{mariani2010industrial, piscitelli2012design, weichslgartner2014daarm, weichslgartner2018design}, we employ a multi-objective evolutionary algorithm for mapping optimization.

To the best of our knowledge, existing \ac{DSE} proposals on mapping optimization in the context of \ac{HAM} strategies, e.g.,~\cite{khanh2013incorporating, mariani2010industrial, piscitelli2012design, skalistis2017near, singh2013accelerating, weichslgartner2014daarm, weichslgartner2018design, ykman2011linking}, consider a fixed isolation scheme that is applied exclusively to \emph{every} resource (core or tile) of \emph{every} explored mapping of \emph{every} application.
For instance, the \ac{DSE} approaches in~\cite{piscitelli2012design,mariani2010industrial, ykman2011linking} assume a core-reservation isolation scheme.
In~\cite{khanh2013incorporating, singh2013accelerating, weichslgartner2014daarm}, a tile-reservation isolation scheme is assumed.
In~\cite{weichslgartner2018design}, a core-sharing isolation scheme is assumed.
The \ac{DSE} proposed in this paper does not assume a globally-fixed isolation scheme.
Instead, it explores the choices of isolation scheme per allocated resource within each explored mapping.
This provides a fine-grained control over the degree of admitted inter-application interferences (which we analyze using the proposed timing analysis) and renders many mappings with promising quality trade-offs reachable to the \ac{DSE} which are not reachable under a fixed isolation scheme.

%%%%%%%%%%%%%%%%%%%%%%%%%%%%%%%%%%%%%%%%%%%%%%%%%%%%%%
%%%%%%%%%%%%%%%%%%%%%%%%%%%%%%%%%%%%%%%%%%%%%%%%%%%%%% (3) PRELIMINARIES
%%%%%%%%%%%%%%%%%%%%%%%%%%%%%%%%%%%%%%%%%%%%%%%%%%%%%%
\section{Preliminaries}\label{sec:preliminaries}
\subsection{Mapping Optimization Problem Specification} 
Similar to any other optimization problem, the multi-/many-core mapping optimization problem requires a problem model which describes the space of possible solutions and the conditions that must be satisfied by a mapping to be regarded a valid solution.
In this paper, we use the graph-based system model from~\cite{blickle1998system}.
This model represents the mapping optimization problem by a so-called \emph{specification} which describes the entire design space and is used to generate different mapping solutions.
The specification and each mapping generated based on that consist of an \emph{application graph}, an \emph{architecture graph}, and \emph{mapping edges} connecting them, which will be introduced in~\cref{sec:archModel,sec:appModel,sec:mappingEdges}.

\subsubsection{Application Model} \label{sec:appModel}
\begin{figure}[!t]
\centering
\includegraphics[width=0.9\linewidth,keepaspectratio]{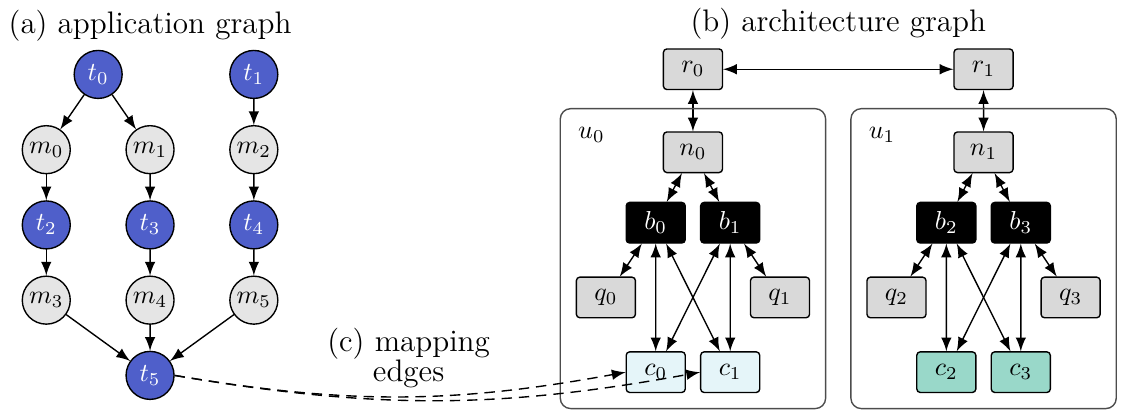}
\caption{Specification of an exemplary mapping optimization problem, composed of (a) application graph, (b) architecture graph, and (c) mapping edges connecting them (depicted only for task $\symTask_5$).}
\label{fig:specification}
\end{figure}

We consider periodic hard real-time applications.
Each application is specified by an application graph ${G_P(\symAllTasks \cup \symAllMessages,\symAppGraphAllEdges)}$ where $\symAllTasks$ denotes the set of (processing) tasks, $\symAllMessages$ denotes the set of messages exchanged between tasks, and edges $\symAppGraphEdge \in \symAppGraphAllEdges$ define data dependencies among tasks and messages, e.g., see~\cref{fig:specification}a.
For each task $\symTask \in \symAllTasks$, the execution period $\symTaskPeriod(\symTask)$, the context-independent worst-case execution time ${\symTaskWCET(\symTask,\symProcessor)}$ on each mappable core $\symProcessor$, and the memory demand $\symTaskMD(\symTask)$ (maximum number of memory accesses) per execution iteration are known.
For each message $\symMessage \in \symAllMessages$, the transfer period $\symMessagePeriod(\symMessage)$ and the maximum payload size $\symMessageSize(\symMessage)$ together with its corresponding memory demand $\symTaskMD(\symMessage)$ (number of memory accesses for reading/writing $\symMessage$ from/to memory) are given.

\subsubsection{Architecture Model} \label{sec:archModel}

We consider heterogeneous tiled many-core architectures composed of multi-core tiles interconnected by a \ac{NOC}.
The platform architecture is specified by an architecture graph ${G_A=(\symAllRouters \cup \symAllNIs \cup \symAllBuses \cup \symAllProcessors \cup \symAllMemories \cup \symAllTiles,\symAllLinks)}$, e.g., see~\cref{fig:specification}b.
Here, $\symRouter \in \symAllRouters$ denotes a \ac{NOC} router, $\symNI \in \symAllNIs$ a \ac{NA}, $\symBus \in \symAllBuses$ a memory bus, $\symProcessor \in \symAllProcessors$ a processing core, and $\symMemory \in \symAllMemories$ a shared memory (or memory bank) with an own memory bus, respectively.
Edges $\symLink \in \symAllLinks$ represent bidirectional connections between these resources.
Each $\symTile \in \symAllTiles$ represents a compute tile which comprises a set of cores and memories and a \ac{NA}, thus, $\symTile \subseteq \symAllNIs \cup \symAllBuses \cup \symAllProcessors \cup \symAllMemories$.
Each memory (or memory bank) is accessible to cores and the \ac{NA} on the same tile via a shared bus.
Each \ac{NA} consists of separate \ac{TX} and \ac{RX} units with own ports to each memory bus.
We assume many-core architectures without timing anomalies~\cite{reineke2006definition}.
This enables a compositional analysis of execution, communication, and memory latency contributions which can be combined to derive globally safe timing guarantees.

\textbf{Memory Model.}
We consider a \ac{NORMA} scheme which is commonly practiced in many-core systems to achieve scalability~\cite{madalozzo2016scalability}.
Under a \ac{NORMA} scheme, the memories located on one tile are not accessible to resources located on other tiles, thus, data exchanges among different tiles are realized exclusively by means of explicit message passing between them.
To achieve storage composability, (a)~the on-tile memory space is partitioned, such that each core is given a dedicated storage space.
Likewise, (b)~shared caches are partitioned or disabled.
In the same line, (c)~the dedicated storage space of each core is dynamically partitioned among the tasks hosted by it.
Finally, (d)~each message is provided with a dedicated memory space in each tile where it is produced and/or consumed.
To perform a memory access, the requestor must first attain the ownership of the shared bus associated with the target memory.
We assume blocking and indivisible memory accesses, so that the processing progress on the requestor side is stalled during the memory operation.
Each memory access is a single-word operation with a known maximum service time, eliminating burst/block operations.
A requestor may perform several consecutive but non-overlapping single-word memory operations during its bus ownership interval.

\textbf{\ac{NOC} Model.}
We consider wormhole-switched \acp{NOC} with a credit-based virtual-channel flow control, for instance~\cite{heisswolf2013providing}, to allow per-link bandwidth reservation and, thereby, to enable link sharing without violating composability.
Under a \emph{wormhole-switched} flow control, data packets are decomposed into control flow digits (flits) of fixed size which are then routed over the \ac{NOC} independent from each other in a pipeline fashion~\cite{ni1993survey}.
The \emph{virtual-channel} flow control~\cite{dally1992virtual} provides multiple buffers per physical link to enable transmission preemption and composable link sharing.
The transmission progress of flits in each buffer of a router is controlled using so-called \emph{credits} which reflect the availability of buffer space at the next router.
This realizes a backpressure mechanism inherently.

\textbf{Communication Model.}
Intra-tile communications, i.e., data exchanges \emph{within} one tile, are realized through a dedicated space in the tile's shared memories.
Inter-tile communications, i.e., data exchanges \emph{between} tiles, are realized by explicit message passing over the \ac{NOC}.
For the latter, after the data is written by the sender into the dedicated memory space, the \ac{TX} reads the data, decomposes it into flits, and injects the flits into the \ac{NOC}.
The flits are then routed over the \ac{NOC} toward the destination tile where the \ac{RX} reconstructs the data from the flits and writes it into a dedicated memory space to be read by the receiver thereafter.

\textbf{Resource Arbitration.}
Any hardware resource, e.g., cores, memory buses or the \ac{NOC}, that can be shared among multiple applications is assumed to have an arbitration policy that is both predictable and composable.
While (a)~\emph{predictability} enables formal worst-case timing analysis of each application, (b)~\emph{composability} ensures that worst-case timing bounds can be derived for each application merely based on the resource budgets reserved for it, without any knowledge about other applications that may run concurrently to it and share resources with it.
This definition necessitates a \emph{contention-free} arbitration policy for each and every resource that may be shared among applications.
Time-Division Multiplexing (TDM) and Weighted Round-Robin (WRR) are well-established predictable and composable arbitration policies which serve as primary candidates for composable many-core systems~\cite{goossens2013virtual,hansson2009compsoc,heisswolf2013providing}.
Both \ac{TDM} and \ac{WRR} establish temporal isolation using time-triggered preemption, dividing the access to a resource into time slots of equal length that are periodically assigned to requestors which have reserved one or more slots on that resource.
\ac{TDM} is not work-conserving\footnote{In a work-conserving arbitration policy, time slots are assigned only to requestors with pending requests while idle slots, i.e., those without an access request, are skipped. This scheme results in a varying arbitration period and varying position of assigned slots for each requestor within one arbitration period.} which leads to a poor average-case performance, making it an unattractive candidate for, e.g., systems hosting both real-time and best-effort applications~\cite{kelter2013evaluation}.
Contrarily, \ac{WRR} provides predictability and composability in a work-conserving fashion by skipping over idle slots.
This enables a notable average-case performance improvement in favor of best-effort applications while allowing worst-case timing guarantees to be derived for real-time applications.
We assume each shared resource has a preemptive time-triggered arbitration policy that follows similar principles as \ac{TDM} and \ac{WRR}.

\subsubsection{Mapping Edges} \label{sec:mappingEdges}
In the specification of the mapping optimization problem, the application graph and the architecture graph are connected by mapping edges $\symAllMappings \subseteq \symAllTasks \times \symAllProcessors$, e.g., see~\cref{fig:specification}c.
Each mapping edge ${\symMapping = (\symTask, \symProcessor) \in \symAllMappings}$ indicates that task $\symTask \in \symAllTasks$ can be executed on core $\symProcessor \in \symAllProcessors$.

\subsection{Arbitration Tuple}
\label{sec:tuple}

\begin{figure}[!t]
\centering
\includegraphics[width=0.9\linewidth,keepaspectratio]{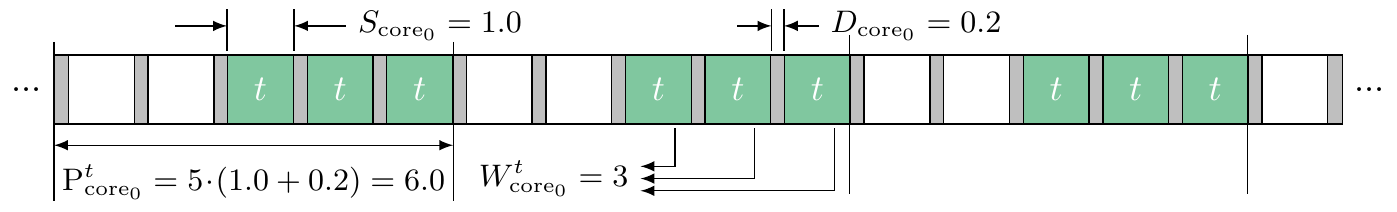}
\caption{Arbitration tuple of task $\symTask$ on core\textsubscript{0} from \cref{fig:motivationIndividualSchemes}a, calculated as $(1.0,3,6.0)$.}
\label{fig:wrr}
\end{figure}

In this work, all parameters relevant for the worst-case timing analysis of a requestor on a resource, i.e., the budget reserved for it and the worst-case interference that
can be imposed by other requestors, are compactly reflected by a so-called \emph{arbitration tuple}: 
For each requestor $x$ of resource $r$, the arbitration tuple is represented as ${(\symArbInterval_r, \symArbWeight^x_r, \symArbPeriod^x_r)}$ where $\symArbInterval_r$ denotes the length of one arbitration slot of $r$, $\symArbWeight_r^x$ denotes the number of periodic arbitration slots (weight) reserved for $x$ on $r$, and $\symArbPeriod^x_r$ denotes the worst-case arbitration period perceived by $x$ on $r$.
Given the arbitration tuple, one can deduce that $x$ has a periodic reserved time budget of ${(\symArbWeight^x_r \cdot \symArbInterval_r)}$ on $r$ and, thus, experiences a worst-case wait time of ${(\symArbPeriod^x_r\!-\!\symArbWeight^x_r \cdot \symArbInterval_r)}$ per arbitration period $\symArbPeriod^x_r$.
We exemplify the calculation of the arbitration tuple for task $\symTask$ in \cref{fig:motivationIndividualSchemes}a under an arbitration policy with a slot length of ${\symArbInterval_\text{core\textsubscript{0}}=1.0}$, an arbitration delay of ${\symArbDelay_\text{core\textsubscript{0}}=0.2}$ between consecutive slots\footnote{The arbitration delay denotes the latency of the arbiter for switching between consecutive arbitration slots. For instance, on a core, it corresponds to the context-switch overhead of the operating system.}, and an arbitration capacity of ${\symArbCapacity_\text{core\textsubscript{0}}=5}$ slots per period.
Under such an arbitration policy, task $\symTask$ with 3 reserved periodic slots perceives an arbitration weight of ${\symArbWeight^\symTask_\text{core\textsubscript{0}}=3}$ and a worst-case arbitration period of ${\symArbPeriod^\symTask_\text{core\textsubscript{0}}=5 \times (1.0+0.2)=6.0}$ based on which the arbitration tuple is created as ${(1.0, 3, 6.0)}$.
This calculation is also illustrated in \cref{fig:wrr}.
Note that, the arbiter delay is reflected in the calculation of arbitration period, rendering the arbitration tuple expressive of realistic resource arbiters in practice.

It may happen that the isolation scheme of a resource leads to one or more arbitration slots to be allocated by the mapping under analysis but not utilized.
Given a work-conserving arbitration policy, e.g.~\ac{WRR}, in such cases we reduce the arbitration capacity of that resource to exclude those slots that are never utilized and, hence, are always skipped by the arbiter. 
For instance, in~\cref{fig:motivationIndividualSchemes}b, core\textsubscript{0} is allocated exclusively where only 3 slots are utilized by $\symTask$ while the remaining 2 slots are never utilized.
Thus, given a work-conserving arbitration policy, the arbitration period is guaranteed not to exceed ${\symArbPeriod^\symTask_\text{core\textsubscript{0}}=(5-2) \times (1.0+0.2)=3.6}$, resulting in an adapted arbitration tuple of ${(1.0, 3, 3.6)}$ for $\symTask$ on core\textsubscript{0}.

The proposed timing analysis presented in~\cref{sec:timingAnalysis} takes the arbitration tuples as input and derives safe bounds on the worst-case timing characteristics of each task and message under the combination of applied isolation schemes.
For the calculation of the arbitration tuples, the arbitration policy of each resource $r$ and its parameters, namely, arbitration slot length $\symArbInterval_r$, arbitration delay $\symArbDelay_r$, and arbitration capacity ${\symArbCapacity_r}$, are provided by the architecture, while both the isolation scheme of each resource $r$ and the number $\symArbWeight_r^x$ of slots reserved for each analyzed requestor $x$ are decided by the mapping optimizer during the \ac{DSE}.
We calculate arbitration tuples for the following requestors:
\begin{itemize}
 \item For each task $\symTask \in \symAllTasks$, the arbitration tuple on core $\symProcessor \in \symAllProcessors$ executing $\symTask$ is calculated and denoted as ${(\symArbInterval_\symProcessor, \symArbWeight^\symTask_\symProcessor, \symArbPeriod^\symTask_\symProcessor)}$.
 \item For each inter-tile message $\symMessage \in \symAllMessages$, the arbitration tuple (a)~on the \ac{TX} $\symTX$ injecting $\symMessage$ into the \ac{NOC}, (b)~on the \ac{NOC} route $\rho$ (sequence of links) over which $\symMessage$ is routed, and (c)~on the \ac{RX} $\symRX$ receiving $\symMessage$ from the \ac{NOC} are calculated and denoted as $(\symArbInterval_\symTX, \symArbWeight^m_\symTX, \symArbPeriod^\symMessage_\symTX)$, ${(\symArbInterval_\rho,\symArbWeight^\symMessage_\rho,\symArbPeriod^\symMessage_\rho)}$, and $(\symArbInterval_\symRX, \symArbWeight^m_\symRX, \symArbPeriod^\symMessage_\symRX)$, respectively.
 \item For each core $\symProcessor \in \symAllProcessors$ that has at least one task mapped to it, the arbitration tuple on each on-tile memory bus $\symBus \in \symAllBuses$ is calculated and denoted as ${(\symArbInterval_\symBus, \symArbWeight^\symProcessor_\symBus, \symArbPeriod^\symProcessor_\symBus)}$.
 \item For each \ac{TX} $\symTX$ that routes at least one outbound message out of the tile, the arbitration tuple on each on-tile memory bus $\symBus \in \symAllBuses$ is calculated and denoted as ${(\symArbInterval_\symBus, \symArbWeight^\symTX_\symBus, \symArbPeriod^\symTX_\symBus)}$.
 \item For each \ac{RX} $\symRX$ that routes at least one incoming message into the tile, the arbitration tuple on each on-tile memory bus $\symBus \in \symAllBuses$ is calculated and denoted as ${(\symArbInterval_\symBus, \symArbWeight^\symRX_\symBus, \symArbPeriod^\symRX_\symBus)}$.
\end{itemize}

%%%%%%%%%%%%%%%%%%%%%%%%%%%%%%%%%%%%%%%%%%%%%%%%%%%%%%
%%%%%%%%%%%%%%%%%%%%%%%%%%%%%%%%%%%%%%%%%%%%%%%%%%%%%% (4) PROPOSED DSE
%%%%%%%%%%%%%%%%%%%%%%%%%%%%%%%%%%%%%%%%%%%%%%%%%%%%%%

\section{Isolation-Aware Design Space Exploration (DSE)}
\label{sec:dse}
\begin{figure}[!t]
\centering
\includegraphics[width=0.75\linewidth,keepaspectratio]{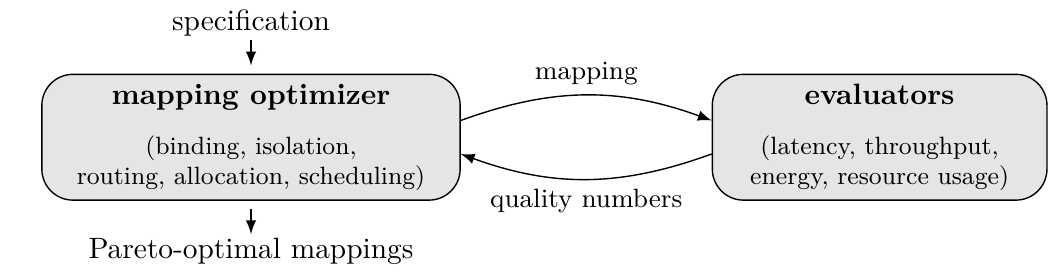}
\caption{Overview of the proposed isolation-aware design space exploration (DSE) approach. 
}
\label{fig:dse}
\end{figure}

This section presents our isolation-aware \ac{DSE} approach, illustrated in~\cref{fig:dse}.
The \ac{DSE} takes as input the specification of the mapping optimization problem comprising the application graph, the architecture graph, and mapping edges, and delivers a set of mappings that offer Pareto-optimal trade-offs w.r.t.\ a given set of design objectives, e.g., latency, throughput, energy, and resource usage.
The \ac{DSE} employs a \emph{mapping optimizer} to explore the space of possible mappings of the application on the target architecture.
The optimizer creates each mapping by conducting a series of binding, isolation, routing, allocation, and scheduling design decisions, elaborated later in~\cref{sec:mappingCreation}.
Once a mapping is generated, it is provided to a set of \emph{evaluators} to assess the quality of the mapping w.r.t.\ the given design objectives.
The proposed isolation-aware timing analysis, which will be presented in~\cref{sec:timingAnalysis}, is used here as an evaluator to derive safe bounds for timing-related design objectives, namely, latency and throughput.
The optimizer uses an evolutionary algorithm which conducts several iterations to generate new mappings and collect a set of Pareto-optimal mappings.

The Pareto-optimal mappings will be used at run time to launch the application on demand by selecting a mapping that complies with the current real-time constraints of the application and resource availability of the system (restricted due to, e.g., other running applications).
Since our timing analysis accounts for the worst-case interferences that may be imposed by any mix of (statically unknown) concurrent applications, the worst-case timing characteristics it provides for each mapping are guaranteed to hold regardless of the mix and deployment of other applications at run time.
This allows the \ac{DSE} to be performed for each application \emph{individually} without any knowledge about the other applications in the system.

\subsection{Mapping Creation}
\label{sec:mappingCreation}
The creation of each mapping starts by making the \emph{binding} and the \emph{isolation} design decisions:
\begin{itemize}
 \item \textbf{Binding.} In this step, each task of the application is bound to a core on the architecture.
	We use the SAT-Decoding approach~\cite{sat_decoding} to explore the binding of tasks to cores.
	By encoding the constraints from \cite{martins_routing_constraints}, we ensure that each task is bound exactly to one core.
 \item \textbf{Isolation.} In this step, an isolation scheme is selected for each core and each tile.
        To this end, each core and each tile is decided to be either \emph{shared} or \emph{reserved}.
        We implement the exploration of isolation schemes using the SAT-Decoding approach~\cite{sat_decoding}, see also~\cite{glass2017handbook}.
\end{itemize}
The creation of the mapping is completed by deriving the implications of binding and isolation decisions on message \emph{routing}, resource \emph{allocation}, and \emph{scheduling} of tasks and message:
\begin{itemize}
 \item \textbf{Routing.} 
	In this step, for each inter-tile message (i.e., a message communicated between two tasks which, according to the binding decisions, are bound to different tiles), a \ac{NOC} route (sequence of links) is determined over which the message is transferred between the two tiles.
	Without loss of generality, we use the XY-routing algorithm~\cite{ni1993survey}.
 \item \textbf{Allocation.} 
	In this step, the processor budget required for the mapping is allocated according to the previously made binding and isolation decisions:
	If a task is bound onto any core of a reserved tile, the whole tile is allocated completely and, thus, none of the resources on that tile can be used by other applications (tile reservation).
	If a task is bound to a reserved core on a shared tile, that particular core is allocated completely while other resources on the tile can be used by other applications (core reservation).
	Finally, if a task is bound to a shared core on a shared tile, only the minimum budget required by that task will be allocated on that core while the remaining core budget can be used by other applications (core sharing).
	Any core and tile that does not correspond to one of the cases above is not allocated and, therefore, can be used by other applications.
	We also allocate the minimum budget required for each inter-tile message on \ac{NOC} links that are part of the message's route specified in the routing step.
 \item \textbf{Scheduling.} We elaborate on the scheduling step in~\cref{sec:scheduling}.
\end{itemize}

\subsection{Isolation-Aware Scheduling}
\label{sec:scheduling}
In the scheduling step, the arbitration tuples listed in~\cref{sec:tuple} which are required for the timing analysis of the mapping are calculated.
In what follows, we first calculate in~\cref{sec:schedulingBudgets} the resource budgets (arbitration weight) required for tasks and messages.
Based on these weights, the arbitration tuples are then calculated in~\cref{sec:schedulingArbitrationTuples}.

\subsubsection{Resource Budget Calculation}
\label{sec:schedulingBudgets}
To calculate the arbitration weight of each task/message $x \in \symAllTasks \cup \symAllMessages$ on each respective resource $r$, first the worst-case arbitration period $\symArbPeriod^x_r$ for $x$ on $r$ is calculated (as presented in~\cref{sec:tuple}) based on $r$'s arbitration capacity $\symArbCapacity_r$ and arbitration slot length $\symArbInterval_r$ which are provided by the architecture.
Then, for each task $\symTask$ bound to core $\symProcessor$, we start with an arbitration weight of $\symArbWeight^\symTask_\symProcessor=1$ and iteratively (I)~construct the arbitration tuple ${(\symArbInterval_\symProcessor, \symArbWeight^\symTask_\symProcessor, \symArbPeriod^\symTask_\symProcessor)}$ and (II)~use \cref{eq:responseTime,eq:nBus,eq:busInterference,eq:coreInterference} from~\cref{sec:timingAnalysis} to determine $\symTask$'s \ac{WCRT} for the current arbitration weight $\symArbWeight^\symTask_\symProcessor$.
If the \ac{WCRT} of $\symTask$ exceeds its deadline given by the application, i.e., ${\text{WCRT}(\symTask) > \symTaskPeriod(\symTask)}$, we (III)~increment $\symArbWeight^\symTask_\symProcessor$ by one and go back to step~(I).
Otherwise, the iterations are terminated and the current arbitration weight is considered for $\symTask$.
Likewise, for each inter-tile message $\symMessage$ routed via \ac{TX} unit $\symTX$, \ac{RX} unit $\symRX$, and \ac{NOC} route $\rho$, we use \cref{eq:wcrl,eq:nMessage,eq:txInterference,eq:RoutingInterference} from~\cref{sec:timingAnalysis} to calculate the minimum arbitration weight ${\symArbWeight^\symMessage_\symTX = \symArbWeight^\symMessage_\symRX=\symArbWeight^\symMessage_\rho}$ such that $\symMessage$'s \ac{WCTT} does not exceed its production period, i.e., ${\text{WCTT}(\symMessage) \leq \symMessagePeriod(\symMessage)}$.

After the weights are derived, we evaluate the overall weight demanded on each resource.
If the weight demanded by all tasks mapped to a core exceed the core's arbitration capacity, the mapping is considered as infeasible and is discarded.
Likewise, if the weight demanded on a \ac{NOC} link, a \ac{TX} or a \ac{RX} by messages routed over it exceeds its arbitration capacity, the mapping is considered as infeasible and discarded.

\subsubsection{Arbitration Tuple Calculation}
\label{sec:schedulingArbitrationTuples}
Given the arbitration weights of tasks and messages on their respective resources, calculated in~\cref{sec:schedulingBudgets}, and the isolation scheme of each core/tile, the arbitration tuples for tasks and messages are calculated as follows.
For each exclusively allocated core $\symProcessor$, first the arbitration capacity is reduced as ${\symArbCapacity_\symProcessor = \sum\nolimits_{\symTask \in \symAllTasks}{\symArbWeight^\symTask_\symProcessor}}$ to discard scheduling slots that are not utilized by the tasks mapped to it, and the arbitration period $\symArbPeriod^\symTask_\symProcessor$ of each task $\symTask \in \symAllTasks$ mapped to it is refined accordingly as presented in~\cref{sec:tuple}.
Then, the arbitration tuple ${(\symArbInterval_\symProcessor, \symArbWeight^\symTask_\symProcessor, \symArbPeriod^\symTask_\symProcessor)}$ is created for each task $\symTask \in \symAllTasks$ mapped to $\symProcessor$ where the scheduling slot length $\symArbInterval_\symProcessor$ is provided by the architecture.
Likewise, for each exclusively allocated tile, first the arbitration capacity of the \ac{TX} unit $\symTX$ is reduced as ${\symArbCapacity_\symTX = \sum\nolimits_{\symMessage \in \symAllMessages}{\symArbWeight^\symMessage_\symTX}}$ to reflect only the arbitration slots reserved for the outbound messages $\symMessage \in \symAllMessages$ of that tile within the current mapping.
Then, the arbitration period $\symArbPeriod^\symMessage_\symTX$ of each outbound message $\symMessage$ on $\symTX$ is refined, and the arbitration tuple $(\symArbInterval_\symTX, \symArbWeight^m_\symTX, \symArbPeriod^\symMessage_\symTX)$ is constructed.
A similar procedure is followed for the incoming messages of the tile.
Since the \ac{NOC} is assumed to be shared, no reduction will be applied to the capacity of \ac{NOC} links, and thus, no refinement is required for the calculation of the arbitration tuple ${(\symArbInterval_\rho,\symArbWeight^\symMessage_\rho,\symArbPeriod^\symMessage_\rho)}$ of an inter-tile message $\symMessage \in \symAllMessages$ on the links of its \ac{NOC} route $\rho$.

For each core ${\symProcessor \in \symAllProcessors}$, the arbitration tuple ${(\symArbInterval_\symBus, \symArbWeight^\symProcessor_\symBus, \symArbPeriod^\symProcessor_\symBus)}$ on each memory bus ${\symBus \in \symAllBuses}$ connected to it is calculated as follows. The bus arbitration slot length $\symArbInterval_\symBus$, the bus arbitration capacity $\symArbCapacity_\symBus$, and the core arbitration weight $\symArbWeight^\symProcessor_\symBus$ on the bus are provided by the architecture.
The arbitration period $\symArbPeriod^\symProcessor_\symBus$ perceived by $\symProcessor$ on bus $\symBus$ is calculated according to the decided isolation scheme: 
If a tile-reservation isolation scheme is selected, the bus arbitration capacity $\symArbCapacity_\symBus$ is reduced to exclude the arbitration slots corresponding to idle cores (cores that do not host any tasks).
Accordingly, the refined arbitration period $\symArbPeriod^\symProcessor_\symBus$ is calculated as presented in~\cref{sec:tuple}.
For each \ac{TX} and \ac{RX}, the arbitration tuples ${(\symArbInterval_\symBus, \symArbWeight^\symTX_\symBus, \symArbPeriod^\symTX_\symBus)}$ and ${(\symArbInterval_\symBus, \symArbWeight^\symRX_\symBus, \symArbPeriod^\symRX_\symBus)}$ on each memory bus ${\symBus \in \symAllBuses}$ connected to them is calculated similarly.
Note that all arbitration capacity reductions discussed above are applied only in case of a work-conserving arbitration policy, e.g., \ac{WRR}.
Otherwise, the arbitration capacities remain unaffected.

%%%%%%%%%%%%%%%%%%%%%%%%%%%%%%%%%%%%%%%%%%%%%%%%%%%%%%
%%%%%%%%%%%%%%%%%%%%%%%%%%%%%%%%%%%%%%%%%%%%%%%%%%%%%% (5) PROPOSED TIMING ANALYSIS
%%%%%%%%%%%%%%%%%%%%%%%%%%%%%%%%%%%%%%%%%%%%%%%%%%%%%%
\section{Isolation-Aware Timing Analysis}
\label{sec:timingAnalysis}
This section presents the proposed timing analysis which formally bounds the \ac{WCRT} of each task $\symTask \in \symAllTasks$ and the \ac{WCTT} of each inter-tile message $\symMessage \in \symAllMessages$ using the arbitration tuples calculated in the scheduling step.
The timing analysis is performed compositionally~\cite{hahn2015towards}, so that the worst-case timing behavior of each task/message is decomposed into several timing contributions on different resources which are analyzed separately and, then, are combined to obtain globally safe timing guarantees.
The obtained task/message latency bounds are then used to bound the worst-case latency (makespan) and throughput of the mapping.

\subsection{Worst-Case Response Time}
The \ac{WCRT} of a task denotes the worst-case time interval between its start time and completion time, subject to the worst-case timing interferences that may arise on shared resources due to the presence of interfering requests.
In each iteration of its execution, a task reads its required data and input messages from the memory, performs its processing, and writes its output messages into the memory.
In general, two sources of interference may occur here:~(a)~bus interferences when accessing memory and (b)~preemption delays on the cores.
Therefore, the \ac{WCRT} of each task $\symTask \in \symAllTasks$ can be bounded as:
\begin{equation}
\text{WCRT}(\symTask,\symProcessor,\symMemory,\symBus) = \symTaskWCET(\symTask,\symProcessor) + \symTaskMD(\symTask) \cdot \symMemDelay(\symMemory, \symBus) + I^{\text{bus}}(\symTask,\symProcessor,\symMemory, \symBus) + I^{\text{core}}(\symTask,\symProcessor,\symMemory,\symBus) 
\label{eq:responseTime}
\end{equation}
where ${\symTaskWCET(\symTask,\symProcessor)}$ denotes the worst-case execution time of $\symTask$ on core $\symProcessor$ in isolation, assuming a zero delay for bus and memory.
$\symTaskMD(\symTask)$ denotes the memory demand (number of single-word memory accesses) of $\symTask$, and $\symMemDelay(\symMemory,\symBus)$ denotes the service time of memory $\symMemory$ over bus $\symBus$, i.e., the turnaround delay of a single-word memory access in absence of bus interferences.
$I^{\text{bus}}(\symTask,\symProcessor,\symMemory,\symBus)$ denotes the worst-case delay introduced due to interferences on bus $\symBus$ over which memory $\symMemory$ is accessed.
$I^{\text{core}}(\symTask,\symProcessor,\symMemory,\symBus)$ denotes the worst-case preemption delay imposed on $\symTask$'s execution on $\symProcessor$.
Here, ${\symTaskWCET(\symTask,\symProcessor)}$ and $\symTaskMD(\symTask)$ are provided by the application, $\symMemDelay(\symMemory,\symBus)$ is provided by the architecture, and $I^{\text{bus}}(\symTask,\symProcessor,\symMemory,\symBus)$ and $I^{\text{core}}(\symTask,\symProcessor,\symMemory,\symBus)$ are derived as follows.

\subsubsection{Memory Bus Interference} 
Given that task $\symTask$ executed on core $\symProcessor$ accesses memory $\symMemory$ over bus $\symBus$, we bound the worst-case memory bus interference $I^{\text{bus}}(\symTask,\symProcessor,\symMemory,\symBus)$ based on the memory demand $\symTaskMD(\symTask)$ of $\symTask$, the service time $\symMemDelay(\symMemory,\symBus)$ of memory $\symMemory$ accessed over bus $\symBus$, and the arbitration tuple ${(\symArbInterval_\symBus, \symArbWeight^\symProcessor_\symBus, \symArbPeriod^\symProcessor_\symBus)}$ for core $\symProcessor$ on bus $\symBus$.
To this end, we first derive the maximum number of dedicated bus arbitration slots, denoted by $N(\symTask,\symProcessor,\symMemory,\symBus)$, that $\symTask$ may require for performing $\symTaskMD(\symTask)$ memory operations:
On the one hand, $N(\symTask,\symProcessor,\symMemory,\symBus)$ can never exceed $\symTaskMD(\symTask)$, since each bus arbitration slot is necessarily long enough to allow performing at least one single-word memory access, i.e., ${\symArbInterval_\symBus \geq \symMemDelay(\symMemory, \symBus)}$.
On the other hand, $N(\symTask,\symProcessor,\symMemory,\symBus)$ can never exceed the maximum number of bus arbitration slots that may pass during $\symTask$'s execution when performing its memory accesses in absence of bus interferences.
Thus, $N(\symTask,\symProcessor,\symMemory,\symBus)$ can be bounded as:
\begin{equation}
 N(\symTask,\symProcessor,\symMemory,\symBus)= \min{ \left\{ \symTaskMD(\symTask), \left\lceil  \dfrac{\symTaskWCET(\symTask,\symProcessor) + \symTaskMD(\symTask) \cdot \symMemDelay(\symMemory, \symBus)}{\symArbInterval_\symBus} \right\rceil \right\}}
 \label{eq:nBus}
\end{equation}
In the worst case, the memory accesses of $\symTask$ are distributed such that each of the $N(\symTask,\symProcessor,\symMemory,\symBus)$ required bus slots falls into a separate bus arbitration period.
In each bus arbitration period, a worst-case wait time of $(\symArbPeriod^\symProcessor_\symBus - \symArbWeight^\symProcessor_\symBus \cdot \symArbInterval_\symBus)$ may be imposed on $\symTask$'s execution before $\symTask$ acquires the bus ownership. 
Therefore, the overall bus interference can be bounded as:
\begin{equation}
I^{\text{bus}}(\symTask,\symProcessor,\symMemory,\symBus) = N(\symTask,\symProcessor,\symMemory,\symBus) \cdot \left(\symArbPeriod^\symProcessor_\symBus - \symArbWeight^\symProcessor_\symBus \cdot \symArbInterval_\symBus \right)
\label{eq:busInterference}
\end{equation}
In some cases, a memory access may be initiated so late that it cannot be completed before the end of the respective bus arbitration slot.
To compensate for this misalignment, the arbitration delay $\symArbDelay_\symBus$ of each bus $\symBus \in \symAllBuses$ is extended by the service time $\symMemDelay(\symMemory,\symBus)$ of its memory $\symMemory$.
This practice (a)~allows late memory accesses to be completed before the next bus arbitration slot is assigned and, thereby, (b)~eliminates memory interferences due to overlapping memory accesses from different bus masters in consecutive bus slots.
Note that, if $\symTask$ accesses multiple memories, the analysis above must be applied for each memory bus $\symBus_i$ over which $\symTask$ performs $\symTaskMD_{i}(\symTask)$ accesses to memory $\symMemory_i$ with a service time of $\symMemDelay(\symMemory_i, \symBus_i)$.

\subsubsection{Core Preemption Delay}
We bound the worst-case preemption delay of task $\symTask$ on core $\symProcessor$ based on its arbitration tuple on $\symProcessor$, i.e., ${(\symArbInterval_\symProcessor, \symArbWeight^\symTask_\symProcessor, \symArbPeriod^\symTask_\symProcessor)}$, and the arbitration tuple of $\symProcessor$ on the bus $\symBus$ over which $\symProcessor$ accesses memory $\symMemory$, i.e., ${(\symArbInterval_\symBus, \symArbWeight^\symProcessor_\symBus, \symArbPeriod^\symProcessor_\symBus)}$.
In each iteration of its execution, $\symTask$ requires an overall dedicated processor time of ${(\symTaskWCET(\symTask,\symProcessor) + \symTaskMD(\symTask) \cdot \symMemDelay(\symMemory) + I^{\text{bus}}(\symTask,\symProcessor,\symMemory,\symBus))}$ on core $\symProcessor$ to complete its processing and perform its memory accesses over the shared bus.
In each scheduling period of $\symProcessor$, an overall processor time of $(\symArbWeight^\symTask_\symProcessor \cdot \symArbInterval_\symProcessor)$ is dedicated to $\symTask$ and, hence, a worst-case periodic wait time of ${(\symArbPeriod^\symTask_\symProcessor- \symArbWeight^\symTask_\symProcessor \cdot {\symArbInterval}_\symProcessor)}$ is imposed on $\symTask$'s execution due to preemption.
Thus, the overall preemption delay imposed on $\symTask$'s execution can be bounded as:
\begin{equation}
I^{\text{core}}(\symTask,\symProcessor,\symMemory,\symBus) = \left\lceil  \dfrac{\symTaskWCET(\symTask,\symProcessor) + \symTaskMD(\symTask) \cdot \symMemDelay(\symMemory) + I^{\text{bus}}(\symTask,\symProcessor,\symMemory,\symBus)}{\symArbWeight^\symTask_\symProcessor \cdot {\symArbInterval}_\symProcessor}  \right\rceil \cdot \left( \symArbPeriod^\symTask_\symProcessor- \symArbWeight^\symTask_\symProcessor \cdot {\symArbInterval}_\symProcessor \right)
\label{eq:coreInterference}
\end{equation}
where the first factor derives the maximum number of scheduling periods over which the execution of $\symTask$ may span, and the second factor reflects the worst-case wait time imposed per scheduling period.
Note that, also here, a memory access may be initiated so late that it cannot be completed before the end of the scheduling slot. 
To compensate for this misalignment, the context-switch latency $\symArbDelay_\symProcessor$ of each core $\symProcessor \in \symAllProcessors$ is extended by the maximum service time among all memories accessible to $\symProcessor$ to allow an initiated memory access to be completed before the following context switch on $\symProcessor$.

\subsection{Worst-Case Traversal Time}
\label{sec:wctt}
The \ac{WCTT} of an inter-tile message $\symMessage$ denotes its worst-case transfer latency from the memory in which $\symMessage$ is stored on the source tile to the respective target memory on $\symMessage$'s destination tile.
This transfer is realized in three steps:
(I)~the \ac{TX} $\symTX$ on the source tile reads $\symMessage$ from memory $\symMemory$ over bus $\symBus$, decomposes it into flits, and injects the flits into the \ac{NOC}.
(II)~the flits are routed over $\symMessage$'s \ac{NOC} route $\rho$ to the destination tile.
Once arrived at the destination tile, (III)~the \ac{RX} $\symRX$ reconstructs $\symMessage$ and writes it into $\symMessage$'s dedicated space in memory $\hat{\symMemory}$ over bus $\hat{\symBus}$.
Therefore, the \ac{WCTT} of message $\symMessage$ can be bounded as:
\begin{equation}
\text{WCTT}(\symMessage,\symTX,\symMemory,\symBus,\rho,\symRX,\hat{\symMemory},\hat{\symBus}) = D^{\text{tx}}(\symMessage,\symTX,\symMemory,\symBus) + D^{\text{noc}}(\symMessage, \rho) + D^{\text{rx}}(\symMessage,\symRX,\hat{\symMemory},\hat{\symBus})
\label{eq:wcrl} 
\end{equation}
where $D^{\text{tx}}(\symMessage,\symTX,\symMemory,\symBus)$, $D^{\text{noc}}(\symMessage, \rho)$, and $D^{\text{rx}}(\symMessage,\symRX,\hat{\symMemory},\hat{\symBus})$ respectively denote the worst-case latency of the transfer steps (I)--(III) above which we analyze in the following.

\subsubsection{TX/RX Latency} 
\label{sec:txLatency}
Reading message $\symMessage$ from memory $\symMemory$ over bus $\symBus$ for injection into the \ac{NOC} is subject to two sources of interference: (a)~\ac{TX} interference (as multiple outbound messages may be transmitted concurrently from the tile) and (b)~bus interference (when \ac{TX} accesses memory for reading $\symMessage$).
We assume that, in each bus arbitration period, bus slots dedicated to \ac{TX} $\symTX$ are used for reading one message only, resulting in an equivalent $\symTX$ slot length of one bus period, i.e., $\symArbInterval_\symTX=\symArbPeriod^\symTX_\symBus$.
To bound the \ac{TX} latency $D^{\text{tx}}(\symMessage,\symTX,\symMemory,\symBus)$, we first derive the maximum number of bus slots, denoted as $N(\symMessage,\symMemory,\symBus)$, that can be required in the worst case for reading $\symMessage$ from memory $\symMemory$:
In each bus arbitration slot, a total of ${\lceil \sfrac{\symArbInterval_\symBus}{\symMemDelay(\symMemory,\symBus)} \rceil}$ consecutive memory accesses can be initiated and completed.
Hence, given $\symMessage$'s memory demand $\symMessageMD(\symMessage)$ (number of memory accesses required for reading $\symMessage$), $N(\symMessage,\symMemory,\symBus)$ is calculated as:
\begin{equation}
N(\symMessage,\symMemory,\symBus) = \left\lceil \symMessageMD(\symMessage) \cdot \left\lceil \dfrac{\symArbInterval_\symBus}{\symMemDelay(\symMemory,\symBus)} \right\rceil^{-1} \right\rceil \label{eq:nMessage}
\end{equation}

Given $N(\symMessage,\symMemory,\symBus)$, we use~\cref{eq:txInterference} to bound the \ac{TX} latency based on the arbitration tuple of $\symTX$ on bus $\symBus$, i.e., $(\symArbInterval_\symBus, \symArbWeight^\symTX_\symBus, \symArbPeriod^\symTX_\symBus)$, and the arbitration tuple of $\symMessage$ on $\symTX$, i.e., $(\symArbInterval_\symTX, \symArbWeight^m_\symTX, \symArbPeriod^\symMessage_\symTX)$ where $\symArbInterval_\symTX=\symArbPeriod^\symTX_\symBus$ as discussed above.
The first summand in~\cref{eq:txInterference} bounds the overall memory service time for the $\symMessageMD(\symMessage)$ memory accesses required for reading $\symMessage$ from memory in absence of bus and \ac{TX} interferences.
The second summand calculates the worst-case bus interference imposed when reading $\symMessage$:
Here, the first factor gives the number of bus arbitration periods that may pass before acquiring all $N(\symMessage,\symMemory,\symBus)$ bus slots required for reading $\symMessage$, and the second factor gives the worst-case wait time imposed per bus period. 
Finally, the third summand in~\cref{eq:txInterference} calculates the worst-case \ac{TX} interference:
Here, the first factor calculates the number of $\symTX$ arbitration periods involved in the transfer of $\symMessage$, and the second factor gives the worst-case wait time per $\symTX$ period.
We use the same analysis to bound the \ac{RX} latency, i.e., $D^{\text{rx}}(\symMessage,\symRX, \hat{\symMemory}, \hat{\symBus})$ in~\cref{eq:wcrl}.
\begin{equation}
\begin{aligned}
D^{\text{tx}}(\symMessage, \symTX, \symMemory, \symBus)  & =  \symMessageMD(\symMessage) \cdot \symMemDelay(\symMemory,\symBus) \;+\; \left\lceil \dfrac{N(\symMessage,\symMemory,\symBus)}{\symArbWeight^{\symTX}_{\symBus}} \right\rceil \cdot \left(\symArbPeriod^\symTX_\symBus-\symArbWeight^\symTX_\symBus \cdot \symArbInterval_\symBus \right)  \\
& + \left\lceil \left\lceil \dfrac{ N(\symMessage,\symMemory,\symBus)}{\symArbWeight^{\symTX}_\symBus} \right\rceil \cdot \dfrac{1}{\symArbWeight^\symMessage_{\symTX}} \right\rceil  \cdot \left(    \symArbPeriod^\symMessage_{\symTX} - \symArbWeight^\symMessage_{\symTX} \cdot \symArbInterval_{\symTX} \right)
\end{aligned}
\label{eq:txInterference}
\end{equation}

\subsubsection{\ac{NOC} Latency}
We bound the worst-case \ac{NOC} routing latency of message $\symMessage$ over route $\rho$ based on the arbitration tuple of $\symMessage$ on the links in $\rho$, i.e., ${(\symArbInterval_\rho,\symArbWeight^m_\rho,\symArbPeriod^m_\rho)}$, using~\cref{eq:RoutingInterference} adopted from~\cite{weichslgartner2014daarm}.
Here, $f_\symMessage$ denotes the number of $\symMessage$'s flits which is calculated based on its payload size $\symMessageSize(\symMessage)$, $|\rho|$ gives the length of route $\rho$ in number of hops, $\tau^{\text{noc}}$ denotes the length of one \ac{NOC} clock cycle which also gives the length of one link arbitration slot ($\symArbInterval_\rho = \symNocClock$), and $D^\text{router}$ gives the latency of a \ac{NOC} router in clock cycles.
The first summand in~\cref{eq:RoutingInterference} derives the transfer latency of $f_m$ flits in absence of interferences on $\rho$.
The second summand bounds the worst-case interferences on $\rho$:
Here, the first factor gives the maximum number of link arbitration periods in which $\symMessage$'s flits may be stalled due to interfering flows, and the second factor gives the worst-case wait time per link arbitration period, see~\cite{weichslgartner2014daarm}.
\begin{equation}
\begin{aligned}
D^{\text{noc}}(\symMessage, \rho) = & \; (f_\symMessage \!-1 + |\rho| \!\cdot\! D^{\text{router}}) \!\cdot\! \symNocClock 
+  \Big( \left\lceil \dfrac{f_\symMessage}{\symArbWeight^\symMessage_\rho} \right\rceil  \!-1 + |\rho| \Big) \!\cdot\! \left( \symArbPeriod^m_\rho - \symArbWeight^m_\rho \!\cdot\! \symNocClock  \right) 
\end{aligned}
\label{eq:RoutingInterference}
\end{equation}

\subsection{Worst-Case Throughput and Latency}
\label{sec:latency}
Given the \ac{WCRT} of each task $\symTask \in \symAllTasks$, in short, $\text{WCRT}(\symTask)$, and the \ac{WCTT} of each inter-tile message $\symMessage  \in \symAllMessages$, in short, $\text{WCTT}(\symMessage)$, the worst-case throughput of the mapping is calculated using~\cref{eq:throughput}.
Likewise, the worst-case application latency (makespan) is calculated using~\cref{eq:makespan} where $\Pi$ denotes the set of all end-to-end paths in the application graph.
\begin{align}
&\mathcal{TH} =  \max\nolimits \Big\{ \max\limits_{\symTask \in \symAllTasks} \left\{{\text{WCRT}(\symTask)}\right\}, \max\limits_{\symMessage \in \symAllMessages} \left\{\text{WCTT}(\symMessage)\right\} \Big\}^{-1} \label{eq:throughput} \\
&\mathcal{L} = \max\limits_{\pi \in \Pi}{\Big\{ \sum\nolimits_{\symTask \in (\symAllTasks \cap \pi)}{\text{WCRT}(\symTask)} + \sum\nolimits_{\symMessage \in (\symAllMessages \cap \pi)}{\text{WCTT}(\symMessage)} \Big\}} \label{eq:makespan}
\end{align}

%%%%%%%%%%%%%%%%%%%%%%%%%%%%%%%%%%%%%%%%%%%%%%%%%%%%%%
%%%%%%%%%%%%%%%%%%%%%%%%%%%%%%%%%%%%%%%%%%%%%%%%%%%%%% (6) EXPERIMENTS
%%%%%%%%%%%%%%%%%%%%%%%%%%%%%%%%%%%%%%%%%%%%%%%%%%%%%%

\section{Experimental Results}
\label{sec:results}
This section presents the results of a series of experiments for a variety of applications and architectures to compare the performance of the proposed isolation-aware approach with existing fixed-isolation-scheme approaches w.r.t.\ the quality of delivered mappings.

\subsection{Experiment Setup} 
\textbf{Applications and Architectures.} 
We use four real-time applications from the domains of networking (7 tasks, 9 messages), consumer (11 tasks, 12 messages), telecommunication (14 tasks, 20 messages), and automotive (18 tasks, 21 messages), provided by the \ac{E3S}~\cite{dick2002embedded}.
For target platform, we use three heterogeneous many-core architectures with $4\!\times\!4$, $5\!\times\!5$, and $6\!\times\!6$ tiles, respectively.
Each architecture is composed of three tile types.
Each tile comprises four homogeneous cores, a shared memory, a \ac{NA} with separate \ac{TX} and \ac{RX} units, and a shared memory bus.
Every shared resource has a \ac{WRR} arbitration policy configured as follows: 
For each core, an arbitration capacity of $\symArbCapacity\!=\!10$, a slot length of $\symArbInterval\!=\!50$\,\textit{us}, and a context switch overhead of $\symArbDelay\!=\!10$\,\textit{us} is considered.
On each bus, each bus master (\ac{TX}, \ac{RX}, and four cores) has an arbitration weight of $\symArbWeight\!=\!1$, resulting in a bus arbitration capacity of $\symArbCapacity\!=\!6$.
The length of each bus slot $\symArbInterval$ is set equal to the memory service time of 7 clock cycles.
For \ac{NOC} links, we consider an arbitration capacity of $\symArbCapacity\!=\!10$ and a slot length $\symArbInterval\!=\symNocClock\!=\!10$\,\textit{ns}.
For each \ac{TX}/\ac{RX}, an arbitration capacity of $\symArbCapacity\!=\!10$ and a slot length equal to the bus arbitration period is considered, cf.~\cref{sec:txLatency}.

\textbf{Design Objectives.} 
We use three design objectives which are commonly considered in the \ac{DSE} of predictable and composable many-core systems:
(I)~\emph{worst-case latency} derived using the proposed timing analysis, 
(II)~\emph{resource usage} calculated (in number of cores) as the sum of time slots reserved on each core divided by core arbitration capacity $K=10$, and (III)~\emph{energy consumption}, calculated based on the processor power parameters provided by~\cite{dick2002embedded} for each investigated benchmark application and the \ac{NOC}/bus energy model from~\cite{wolkotte2005energy}, assuming a link length of $2$\,\textit{mm} and a bus length of $5$\,\textit{mm}.

\textbf{Design Space Exploration.}
To perform the \ac{DSE}, we use the OpenDSE framework~\cite{opendse} and the NSGA-II~\cite{deb2002fast} multi-objective evolutionary algorithm provided by the optimization framework {\sc{Opt4J}}~\cite{lukasiewycz2011opt4j}.
Each run of the \ac{DSE} features 4,000 iterations with 25 mappings generated per iteration and a population size of 100 mappings.
The results reported in this section are an average over 20 runs of the \ac{DSE} for each application on each architecture.

\textbf{Investigated Approaches.} We compare the proposed isolation-aware \ac{DSE} approach with the three major fixed-isolation-scheme \ac{DSE} approaches, namely, core sharing, core reservation, and tile reservation.
For each approach, the \ac{DSE} delivers a set of mappings with Pareto-optimal trade-offs in the space of the three design objectives above.
We compare the \ac{DSE} approaches in terms of the quality of mappings they deliver.

\textbf{Quality Metric.}
To compare the quality of mappings obtained by the investigated \ac{DSE} approaches, we use the well-established \textit{${\epsilon\text{-dominance}}$} metric~\cite{laumanns2002combining} from the domain of multi-objective optimization, defined as follows:
Let $F \subseteq \mathbb{R}^{+^N}$ represent a set of Pareto-optimal mappings $f \in F$ obtained by an optimization approach in the space of design objectives $o_1,...,o_N$ to be minimized. Let $S \subseteq \mathbb{R}^{+^N}$ represent a reference set containing the true Pareto-optimal mappings for the optimization problem in question.
To assess the quality of mappings in $F$ w.r.t.\ $S$, ${\epsilon\text{-dominance}}$ provides a unary indicator $\epsilon_F \in [0,1)$ calculated as:
\begin{equation}
 \epsilon_{F}=\min\{0 \leq \epsilon < 1 \mid  \forall s \in S: \exists f \in F \:\:\text{s.t.}\:\: (1-\epsilon) \!\cdot\! f_{o_n} \leq s_{o_n} \; \forall n=1,...,N \}
\end{equation}
where $f_{o_n}$ and $s_{o_n}$ denote the quality of mappings $f \in F$ and $s \in S$, respectively, w.r.t.\ design objective $o_n$.
A \emph{smaller} value for $\epsilon_F$ indicates a smaller distance between the mappings in $F$ and those in the reference set $S$ w.r.t.\ the design objectives which, in turn, denotes a \emph{higher quality} of mappings in $F$.
For our following experiments, the reference set $S$ is constructed by collecting the mappings obtained by the four approaches under analysis in one set.

\subsection{Result Discussion} 
\begin{figure}[!th]
\centering
\includegraphics[width=1\linewidth,keepaspectratio]{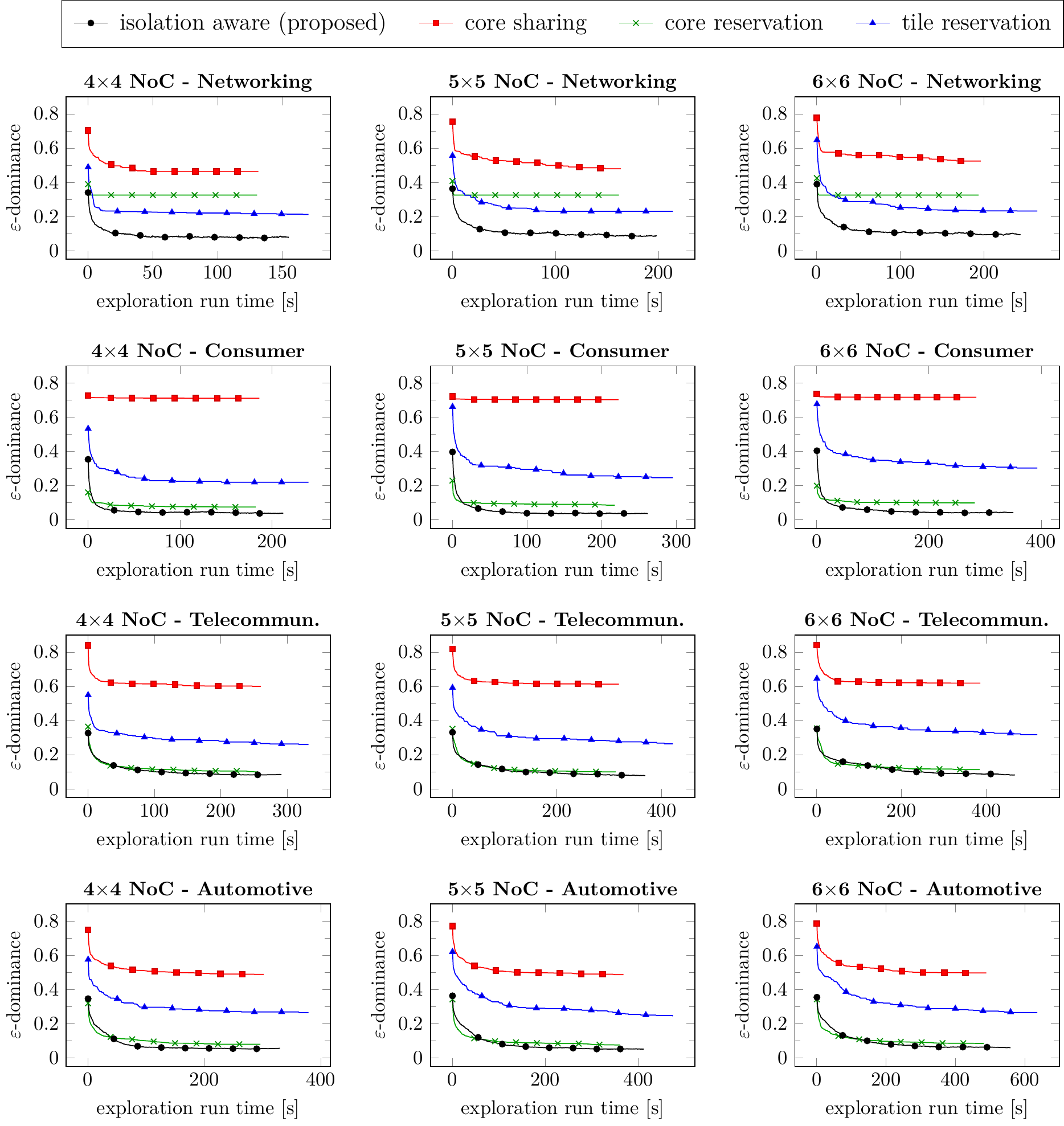}
\caption{${\epsilon\text{-dominance}}$ of the investigated \ac{DSE} approaches versus their exploration run time. Plots in each column (row) correspond to the same many-core architecture (application). The proposed isolation-aware approach outperforms existing fixed-isolation-scheme approaches for all applications and architectures, denoted by its lower ${\epsilon\text{-dominance}}$ indicator.}
\label{fig:epsilon}
\end{figure}

\Cref{fig:epsilon} illustrates, for each benchmark application on each many-core architecture, the ${\epsilon\text{-dominance}}$ indicator for the proposed approach and the three fixed-isolation-scheme approaches across their exploration run time. 
The plots in each row correspond to the same application, and the plots in each column correspond to the same architecture.
In general, the quality of mappings collected by each approach improves as the exploration progresses, resulting in a decrease in the $\epsilon$ indicator of each approach throughout the course of its optimization iterations. 
Also, the exploration time of each approaches grows with the application size (compare the plots in one column) and architecture size (compare the plots in one row).
The results illustrated in~\cref{fig:epsilon} uniformly verify that (a)~the proposed approach always outperforms the approaches with a fixed isolation scheme, irrespective of the choice of application and architecture, which is indicated by its lower ${\epsilon\text{-dominance}}$ indicator at the end of its exploration.
Moreover, (b)~the proposed approach exhibits an exploration run time within the same range as the other approaches and, hence, does not impact the scalability of the optimization approach adversely. % (lowest ${\epsilon\text{-dominance}}$)
Among the other approaches, core reservation generally performs better than core sharing and tile reservation for most of cases. 
Core sharing exhibits the worst quality of solutions for all applications and architectures.
Considering all 12 combinations of applications and architectures, the proposed isolation-aware approach achieves an ${\epsilon\text{-dominance}}$ improvement of up to $67\%$ (compared with the core-sharing approach) with an average improvement of $26\%$ over the three fixed-isolation-scheme approaches.

\begin{figure}[!th]
\centering
\includegraphics[width=0.95\linewidth,keepaspectratio]{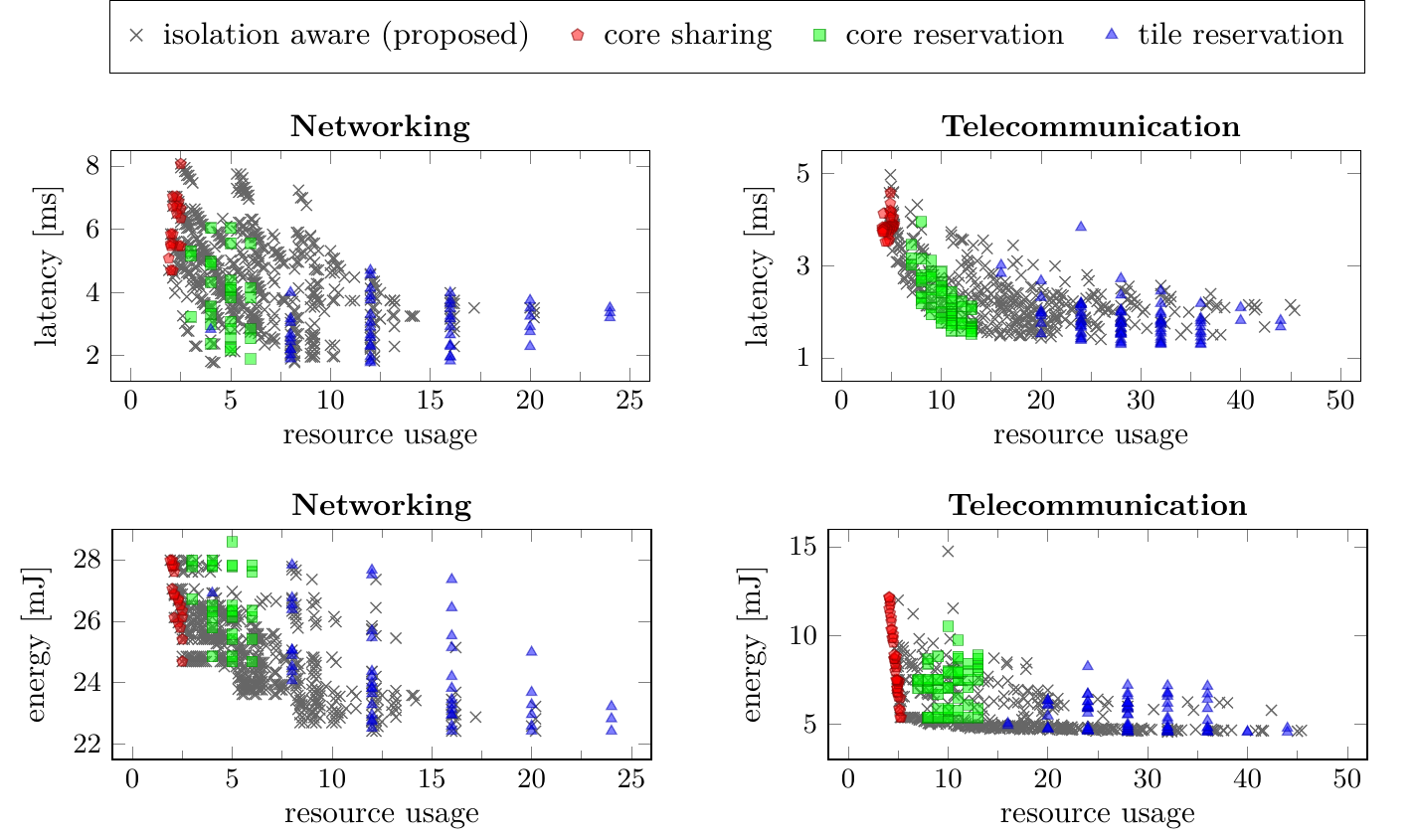}
\caption{2D projections of the 3D objective space showing the spread of solutions delivered by each \ac{DSE} approach for two exemplary applications (columns) on the 6$\boldsymbol\times$6 many-core architecture.}
\label{fig:objectiveSpace}
\end{figure}

To reason about these quality differences, we investigate the distribution of the mappings delivered by each \ac{DSE} approach in the 3D objective space.
\Cref{fig:objectiveSpace}~illustrates this using two 2D projections of the objective space for two exemplary applications on the $6\!\times\!6$ architecture.
Similar distributions are obtained for other applications and architectures.
For each application (column), the x-axes in both projections denote resource usage while the y-axes denote worst-case latency (top) and energy consumption (bottom).
As illustrated, solutions delivered by the core-sharing approach exhibit low resource usage, as this approach allocates a \emph{minimal} resource budget, just enough to meet the deadlines of tasks/messages.
This, however, implies a high worst-case inter-application interference and, thus, considerably high worst-case latency and energy consumption.
The tile-reservation approach, on the other hand, allocates compute tiles exclusively (thus, higher resource usage) which eliminates all on-tile inter-application interferences (thus, lower latency and energy).
When considering all objectives together, both of these approaches need a large scaling factor $\epsilon$ to compensate for their inferior objective values, explaining their poor (high) $\epsilon\text{-dominance}$ indices in~\cref{fig:epsilon}.
A compromise between these approaches is achieved by core reservation which restricts the exclusive allocation of resources to cores, leading to a moderate trade-off between resource usage, latency, and energy consumption and, thus, a better (lower) $\epsilon\text{-dominance}$ index.

The fixed isolation scheme of the approaches above excludes large parts of the actual solution space and, thus, restricts their coverage of the objective space to a considerably smaller sub-region.
Contrary to them, the proposed approach covers and extends beyond the solution space of these approaches as it explores the isolation schemes in combination within each mapping, also reflected by the wide spread of its solutions in~\cref{fig:objectiveSpace}.
As a result, it can find solutions of higher quality and, thus, outperforms the other approaches as confirmed by its lower $\epsilon\text{-dominance}$ index in~\cref{fig:epsilon} for all investigated applications and architectures.

%%%%%%%%%%%%%%%%%%%%%%%%%%%%%%%%%%%%%%%%%%%%%%%%%%%%%%
%%%%%%%%%%%%%%%%%%%%%%%%%%%%%%%%%%%%%%%%%%%%%%%%%%%%%% (7) CONCLUSION
%%%%%%%%%%%%%%%%%%%%%%%%%%%%%%%%%%%%%%%%%%%%%%%%%%%%%%
\section{Conclusion}
\label{sec:conclusion}
\acresetall
Applications in composable many-core systems are typically developed with the assumption of a fixed inter-application isolation scheme which restricts the resource allocation policy of the applications and, therefore, their quality trade-off w.r.t.\ resource usage and worst-case timing.
To lift this restriction, we have proposed (a)~an \emph{isolation-aware \ac{DSE}} which explores the isolation schemes per allocated core/tile within each mapping and (b)~an \emph{isolation-aware timing analysis} to formally bound the worst-case timing properties of each explored mapping.
For a variety of hard real-time applications and many-core architectures, we have experimentally demonstrated the advantage of the proposed approach over existing fixed-isolation-scheme approaches w.r.t.\ the quality of the delivered solutions in terms of resource usage, worst-case latency, and energy consumption.

%%%%%%%%%%%%%%%%%%%%%%%%%%%%%%%%%%%%%%%%%%%%%%%%%%%%%%
%%%%%%%%%%%%%%%%%%%%%%%%%%%%%%%%%%%%%%%%%%%%%%%%%%%%%% 
%%%%%%%%%%%%%%%%%%%%%%%%%%%%%%%%%%%%%%%%%%%%%%%%%%%%%%

\bibliography{references} 

\end{document}